\documentclass{aa}
\usepackage[varg]{txfonts}

\usepackage{graphicx}
\usepackage{color}

\newcommand{\mean}[1]{\left \langle #1 \right \rangle}
\newcommand{\logMhalo}{\log ( M_\mathrm{halo} / h^{-1}\mathrm{M}_\odot )}

\def \speczmeanz {1.19}
\def \speczlogm {10.72}
\def \speczloglxm {33.02}
\def \speczchisq {15.14}
\def \speczdof {7}
\def \speczbias {2.20}
\def \speczdbiaslo {0.45}
\def \speczdbiashi {0.37}
\def \speczmhalo {12.79}
\def \speczdmhalolo {0.43}
\def \speczdmhalohi {0.26}

\def \speczlxmlomeanz {0.88}
\def \speczlxmlologm {10.73}
\def \speczlxmlologlxm {32.53}
\def \speczlxmlochisq {6.91}
\def \speczlxmlodof {6}
\def \speczlxmlobias {2.14}
\def \speczlxmlodbiaslo {0.41}
\def \speczlxmlodbiashi {0.35}
\def \speczlxmlomhalo {13.06}
\def \speczlxmlodmhalolo {0.38}
\def \speczlxmlodmhalohi {0.23}

\def \speczlxmhimeanz {1.50}
\def \speczlxmhilogm {10.73}
\def \speczlxmhiloglxm {33.49}
\def \speczlxmhichisq {1.88}
\def \speczlxmhidof {4}
\def \speczlxmhibias {2.95}
\def \speczlxmhidbiaslo {1.42}
\def \speczlxmhidbiashi {0.93}
\def \speczlxmhimhalo {12.97}
\def \speczlxmhidmhalolo {1.26}
\def \speczlxmhidmhalohi {0.39}

\def \speczmstarlomeanz {0.97}
\def \speczmstarlologm {10.39}
\def \speczmstarlologlxm {33.03}
\def \speczmstarlochisq {7.96}
\def \speczmstarlodof {6}
\def \speczmstarlobias {2.11}
\def \speczmstarlodbiaslo {0.58}
\def \speczmstarlodbiashi {0.45}
\def \speczmstarlomhalo {12.93}
\def \speczmstarlodmhalolo {0.62}
\def \speczmstarlodmhalohi {0.31}

\def \speczmstarhimeanz {1.41}
\def \speczmstarhilogm {11.05}
\def \speczmstarhiloglxm {33.02}
\def \speczmstarhichisq {10.25}
\def \speczmstarhidof {5}
\def \speczmstarhibias {2.69}
\def \speczmstarhidbiaslo {0.79}
\def \speczmstarhidbiashi {0.61}
\def \speczmstarhimhalo {12.90}
\def \speczmstarhidmhalolo {0.62}
\def \speczmstarhidmhalohi {0.30}

\def \speczphotzmeanz {1.20}
\def \speczphotzlogm {10.72}
\def \speczphotzloglxm {33.03}
\def \speczphotzchisq {16.28}
\def \speczphotzdof {7}
\def \speczphotzbias {2.21}
\def \speczphotzdbiaslo {0.40}
\def \speczphotzdbiashi {0.34}
\def \speczphotzmhalo {12.77}
\def \speczphotzdmhalolo {0.37}
\def \speczphotzdmhalohi {0.23}

\def \speczphotzlxmlomeanz {0.88}
\def \speczphotzlxmlologm {10.73}
\def \speczphotzlxmlologlxm {32.53}
\def \speczphotzlxmlochisq {8.45}
\def \speczphotzlxmlodof {5}
\def \speczphotzlxmlobias {2.12}
\def \speczphotzlxmlodbiaslo {0.45}
\def \speczphotzlxmlodbiashi {0.37}
\def \speczphotzlxmlomhalo {13.03}
\def \speczphotzlxmlodmhalolo {0.43}
\def \speczphotzlxmlodmhalohi {0.25}

\def \speczphotzlxmhimeanz {1.52}
\def \speczphotzlxmhilogm {10.73}
\def \speczphotzlxmhiloglxm {33.51}
\def \speczphotzlxmhichisq {1.22}
\def \speczphotzlxmhidof {4}
\def \speczphotzlxmhibias {2.91}
\def \speczphotzlxmhidbiaslo {1.03}
\def \speczphotzlxmhidbiashi {0.75}
\def \speczphotzlxmhimhalo {12.93}
\def \speczphotzlxmhidmhalolo {0.77}
\def \speczphotzlxmhidmhalohi {0.33}

\def \speczphotzmstarlomeanz {0.98}
\def \speczphotzmstarlologm {10.39}
\def \speczphotzmstarlologlxm {33.03}
\def \speczphotzmstarlochisq {9.76}
\def \speczphotzmstarlodof {6}
\def \speczphotzmstarlobias {2.25}
\def \speczphotzmstarlodbiaslo {0.48}
\def \speczphotzmstarlodbiashi {0.40}
\def \speczphotzmstarlomhalo {13.03}
\def \speczphotzmstarlodmhalolo {0.43}
\def \speczphotzmstarlodmhalohi {0.25}

\def \speczphotzmstarhimeanz {1.42}
\def \speczphotzmstarhilogm {11.05}
\def \speczphotzmstarhiloglxm {33.03}
\def \speczphotzmstarhichisq {13.36}
\def \speczphotzmstarhidof {5}
\def \speczphotzmstarhibias {2.18}
\def \speczphotzmstarhidbiaslo {0.78}
\def \speczphotzmstarhidbiashi {0.57}
\def \speczphotzmstarhimhalo {12.53}
\def \speczphotzmstarhidmhalolo {0.98}
\def \speczphotzmstarhidmhalohi {0.39}

\def \specznogroupsmstarlomeanz {0.99}
\def \specznogroupsmstarlologm {10.37}
\def \specznogroupsmstarlologlxm {33.06}
\def \specznogroupsmstarlochisq {6.20}
\def \specznogroupsmstarlodof {5}
\def \specznogroupsmstarlobias {1.69}
\def \specznogroupsmstarlodbiaslo {0.72}
\def \specznogroupsmstarlodbiashi {0.49}
\def \specznogroupsmstarlomhalo {12.50}
\def \specznogroupsmstarlodmhalolo {1.67}
\def \specznogroupsmstarlodmhalohi {0.47}

\def \specznogroupsmstarhimeanz {1.45}
\def \specznogroupsmstarhilogm {11.05}
\def \specznogroupsmstarhiloglxm {33.05}
\def \specznogroupsmstarhichisq {9.12}
\def \specznogroupsmstarhidof {4}
\def \specznogroupsmstarhibias {2.48}
\def \specznogroupsmstarhidbiaslo {0.71}
\def \specznogroupsmstarhidbiashi {0.55}
\def \specznogroupsmstarhimhalo {12.73}
\def \specznogroupsmstarhidmhalolo {0.64}
\def \specznogroupsmstarhidmhalohi {0.32}

\begin{document}

\title{The \textit{XMM}-NEWTON WIDE FIELD SURVEY IN THE COSMOS FIELD: CLUSTERING
DEPENDENCE OF X-RAY SELECTED AGN ON HOST GALAXY PROPERTIES}

\author{A. Viitanen\inst{1}
\and V. Allevato\inst{2,1}
\and A. Finoguenov\inst{1}
\and A. Bongiorno\inst{3}
\and N. Cappelluti\inst{4}
\and R. Gilli\inst{5}
\and T. Miyaji\inst{6}
\and M. Salvato\inst{7}}

\institute{Department of Physics, University of Helsinki, Gustaf Hällströmin
katu 2a, FI-00014 Helsinki, Finland\\
\email{akke.viitanen@helsinki.fi}
\label{inst1}
\and Scuola Normale Superiore, Piazza dei Cavalieri 7, I-56126 Pisa, Italy
\label{inst2}
\and INAF -- Osservatorio Astronomico di Roma, Via Frascati 33, 00078 Monte
Porzio Catone (Roma), Italy
\label{inst3}
\and Physics Department, University of Miami, Knight Physics Building, Coral
Gables, FL 33124, USA
\label{inst4}
\and INAF -- Osservatorio di Astrofisica e Scienza dello Spazio di Bologna, via
Gobetti 93/3, 40129 Bologna, Italy
\label{inst5}
\and Instituto de Astronom\'ia, Universidad Nacional Aut\'onoma de M\'exico,
22860 Ensenada, Mexico
\label{inst6}
\and Max-Planck-Institut für extraterrestrische Physik, Giessenbachstrasse 1,
85748 Garching, Germany
\label{inst7}
}

\date{Received Nameofmonth dd, yyyy; accepted Nameofmonth dd, yyyy}

\abstract
    {}
    {We study the spatial clustering of $632$ $(1130)$ \textit{XMM}-COSMOS
    Active Galactic Nuclei (AGNs) with known spectroscopic (spectroscopic or
    photometric) redshifts in the range $z = [0.1 - 2.5]$ in order to measure
    the AGN bias and estimate the typical mass of the hosting dark matter (DM)
    halo as a function of AGN host galaxy properties. We create AGN subsamples
    in terms of stellar mass $M_*$ and specific black hole accretion rate
    $L_X/M_*$, to probe how AGN environment depends on these quantities.
    Further, we derive the $M_*-M_\mathrm{halo}$ relation for our sample of
    \textit{XMM}-COSMOS AGNs and compare it to results in literature for normal
    non-active galaxies.
    }
    {We measure the projected two-point correlation function $w_p(r_p)$ using
    both the classic and the generalized clustering estimator based on
    photometric redshifts as probability distribution functions in addition to
    any available spectroscopic redshifts. We measure the large-scale ($r_p
    \gtrsim 1 \,{h}^{-1}\mathrm{Mpc}$) linear bias $b$ by comparing the
    clustering signal to that expected of the underlying DM distribution. The
    bias is then related to the typical mass of the hosting halo
    $M_\mathrm{halo}$ of our AGN subsamples. Since $M_*$ and $L_X/M_*$ are
    correlated, we match the distribution in terms of one quantity, while split
    the distribution in the other.}
    {For the full spectroscopic AGN sample, we measure a typical DM halo mass of
    $\logMhalo = \speczmhalo_{-\speczdmhalolo}^{+\speczdmhalohi}$,
    similar to galaxy group environments and in line with previous studies for
    moderate-luminosity X-ray selected AGN. We find no significant dependence on
    specific accretion rate $L_X/M_*$, with
    $\logMhalo = \speczlxmlomhalo_{-\speczlxmlodmhalolo}^{+\speczlxmlodmhalohi}$
    and
    $\logMhalo = \speczlxmhimhalo_{-\speczlxmhidmhalolo}^{+\speczlxmhidmhalohi}$
    for \textit{low} and \textit{high} $L_X/M_*$ subsamples, respectively. We
    also find no difference in the hosting halos in terms of $M_*$ with
    $\logMhalo = \speczmstarlomhalo_{-\speczmstarlodmhalolo}^{+\speczmstarlodmhalohi}$
    (low) and
    $\logMhalo = \speczmstarhimhalo_{-\speczmstarhidmhalolo}^{+\speczmstarhidmhalohi}$
    (high). By comparing the $M_*-M_\mathrm{halo}$ relation derived for
    \textit{XMM}-COSMOS AGN subsamples with what is expected for normal
    non-active galaxies by abundance matching and clustering results, we find
    that the typical DM halo mass of our \textit{high} $M_*$ AGN subsample is
    similar to that of non-active galaxies. However, AGNs in our \textit{low}
    $M_*$ subsample are found in more massive halos than non-active galaxies. By
    excluding AGNs in galaxy groups from the clustering analysis, we find
    evidence that the result for \textit{low} $M_*$ may be due a larger fraction
    of AGNs as satellites in massive halos.}
    {}


\titlerunning{\textit{XMM}-COSMOS AGN Clustering and Host Galaxy Properties}

\maketitle

\section{Introduction}

Supermassive black holes (SMBH) with $M \sim 10^{6-9} \,\mathrm{M}_\odot$ reside
at the centers of virtually every massive galaxy. SMBHs reach these masses by
growing via matter accretion and simultaneously shine luminously as an active
galactic nucleus (AGN). Interestingly, BHs and their host galaxies seem to
co-evolve, as suggested by the correlation between the SMBH and the host galaxy
properties (velocity dispersion, luminosity, stellar mass). However, the
co-evolution scenario, AGN feedback and accretion mechanisms are still poorly
known \citep[e.g.][]{alexander_hickox12}.

AGNs and their host galaxies reside in collapsed dark matter (DM) structures
i.e. halos. In the concordance $\Lambda$CDM cosmology these halos form
hierarchially 'bottom up' from the smallest structures (density fluctuations in
the CMB) that grow via gravitational instability to the largest (galaxy groups
and clusters). AGNs and DM halos they reside in are both biased tracers of the
underlying DM distribution. By measuring the clustering of AGN, and comparing
that to the underlying DM distribution, the AGNs may be linked to their hosting
DM halos \citep[e.g.][]{cappelluti12, krumpe14}. Recent AGN clustering
measurements have not been able to paint a coherent picture of the complex
interplay of AGN and their environment. It seems that optically selected
luminous quasars prefer to live in halos $\mathrm{few} \times 10^{12}
\,{h}^{-1}\mathrm{M}_\odot$ over a wide range in redshift \citep[][]{croom05,
daangela08, ross09} while moderate luminosity X-ray selected AGN prefer larger
halos $10^{12.5-13} \,{h}^{-1}\mathrm{M}_\odot$ at similar redshifts
\citep[][]{coil09, allevato11, koutoulidis13}.

\citet{mendez16} suggest that the clustering of AGN could be understood as the
clustering of galaxies with matched properties in terms of stellar mass and
star-formation rate and redshift, and AGN selection effects. This would indicate
that instead of the properties of the AGN itself, the properties of the host
galaxy, such as, stellar mass $M_*$ or specific black hole accretion rate
$L_X/M_*$ have a more significant role in driving the clustering of AGN.

Many authors have investigated the relation between the stellar mass and the DM
halo mass, the so-called $M_*-M_\mathrm{halo}$ relation, for normal non-active
galaxies via abundance matching \citep[][]{moster13, behroozi13b}, clustering
measurements and HOD modeling \citep[][]{zheng07, wake11} or weak lensing
\citep[][]{coupon15}. For X-ray selected AGNs, the $M_*-M_\mathrm{halo}$
relation has only recently been studied observationally. \citet{georgakakis14}
argue that AGN environment is closely related to $M_*$. However, they do not
measure $M_*$ directly, but use the rest frame absolute magnitude in the J band
as a proxy for $M_*$. Very recently, \citet{mountrichas19} measured the AGN
clustering dependence directly in terms of $M_*$ and found that the environments
of X-ray AGN at $z = 0.6-1.4$ are similar to normal galaxies with matched SFR
and redshift.

In this study, we wish to build upon the previous X-ray selected AGN clustering
measurements in \textit{XMM}-COSMOS \citep[][]{miyaji07, gilli09, allevato11},
to investigate the clustering dependence on host galaxy properties ($M_*$,
$L_X/M_*$). We compare this to the $M_*-M_\mathrm{halo}$ relation for normal
non-active galaxies. In our clustering measurements, we also investigate the new
generalized estimator which has been introduced
\citep[][]{georgakakis14, allevato16}, where photometric redshifts are included
in the clustering analysis as probability density functions. Clustering
measurements using photometric redshifts will be important in future X-ray AGN
surveys, where spectroscopic redshifts are not available either
due to AGN being optically faint, or because no extensive spectroscopic
follow-up campaigns are available. In eROSITA, for example, spectroscopic
redshifts will be available only for a certain portion of the sky, and only at
later stages of the survey \citep{merloni19}.

We adopt a flat $\Lambda$CDM cosmology with $\Omega_m=0.3$,
$\Omega_\Lambda=0.7$, $\sigma_8=0.8$ and $h=0.7$. Distances reported are
comoving distances and the dependence in $h$ is shown explicitly. The symbol
`$\log$` signifies base 10 logarithm. DM halo masses are defined as the enclosed
mass within the Virial radius, within which the mean density is 200 times more
than the background density. DM halo masses scale as $h^{-1}$, while $M_*$
scales as $h^{-2}$.

\section{\textit{XMM}-COSMOS Multiwavelength Data Set}

To study the dependence of AGN clustering in terms of host galaxy properties, we
use the Cosmic Evolution Survey \citep[COSMOS,][]{scoville07}. COSMOS is a
multiwavelength survey over $1.4 \times 1.4 \,\mathrm{deg}^2$ field designed to
study the evolution of galaxies and AGNs up to redshift $z \sim 6$.
To date the field has been covered by a wide variety of
instruments from radio to X-ray bands. \textit{XMM}-Newton surveyed $2.13
\,\mathrm{deg}^2$ of the sky in the COSMOS field in the $0.5-10 \,\mathrm{keV}$
band for a total of $1.55 \,\mathrm{Ms}$ \citep{hasinger07, cappelluti07,
cappelluti09}, providing an unprecedented large sample of point-like X-ray
sources (1822).

\begin{figure}
    \resizebox{\hsize}{!}{\includegraphics[width=\columnwidth]{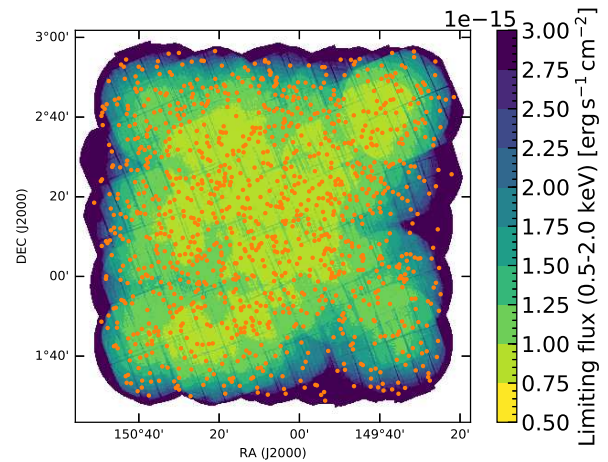}}
    \caption{\textit{XMM}-COSMOS sensitivity map in the soft band $0.5-2.0
    \,\mathrm{keV}$ \citep{cappelluti09}. Orange points mark the positions
    of $1130$ AGN with $z = [0.1-2.5]$ used in this study.}
    \label{fig:radec}
\end{figure}

\citet{brusa10} carried out the optical identification and presented the
multiwavelength properties ($24 \mu\mathrm{m}$ to UV) of ${\sim}1800$ sources
with a spectroscopic completeness of ${\sim} 50\%$ \citep[e.g.][]{hasinger18}.
\citet{salvato09, salvato11} derived accurate photometric redshifts with
$\sigma_{\Delta z / (1 + z_\mathrm{spec})} \sim 0.015$. \citet{bongiorno12} used
a Spectral Energy Distribution (SED) fitting technique based on AGN+Galaxy
template SEDs and estimated the host galaxy properties, i.e. stellar mass $M_*$
and star-formation rate (SFR) of ${\sim} 1700$ AGN in COSMOS up to $z \lesssim
3$. The quantity $L_X/M_*$ corresponds to the rate of accretion onto the central
SMBH scaled relative to the stellar mass of the host galaxy. Assuming a
$M_*-M_\mathrm{BH}$ relation and a constant bolometric correction to convert
from $L_X$ to $L_\mathrm{bol}$, then Eddington ratio ($\lambda_\mathrm{Edd}
\equiv L_\mathrm{bol}/L_\mathrm{Edd}$) can be expressed as:
\begin{equation}
    \lambda_\mathrm{Edd}
    =
    \frac{A \times k_\mathrm{bol}}{1.3 \times 10^{38}} \times \frac{L_X}{M_*}.
\end{equation}
With $A = 500$ and $k_\mathrm{bol} = 25$, $L_X/M_* = 10^{34} \,\mathrm{erg}
\,\mathrm{s}^{-1} \,\mathrm{M}_\sun^{-1}$ corresponds to accretion at Eddington
luminosity i.e. $\lambda_\mathrm{Edd} = 1$ \citep{bongiorno12}.

\begin{figure*}
    \resizebox{\hsize}{!}{
        \includegraphics[width=0.45\textwidth]{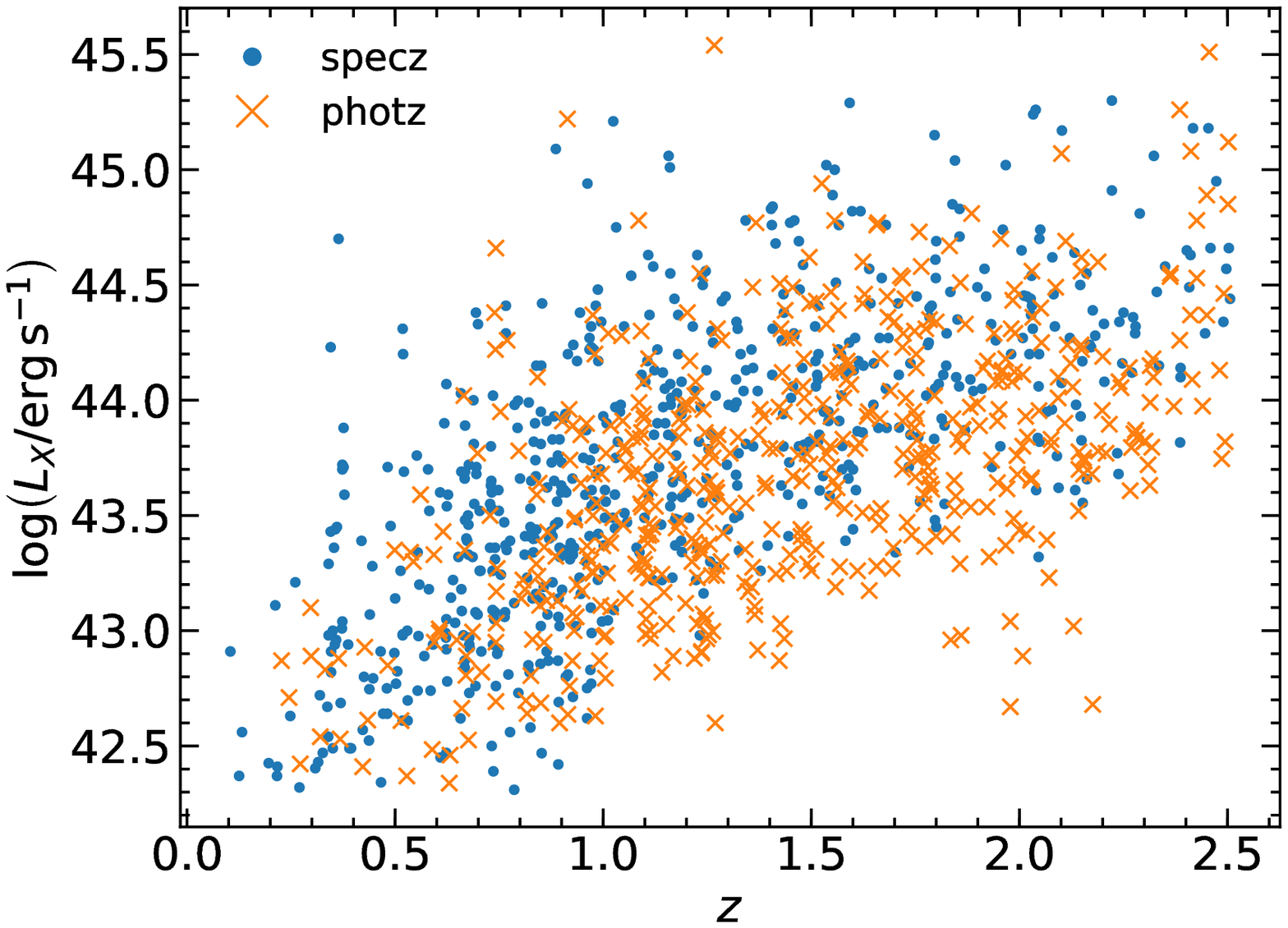}
        \includegraphics[width=0.45\textwidth]{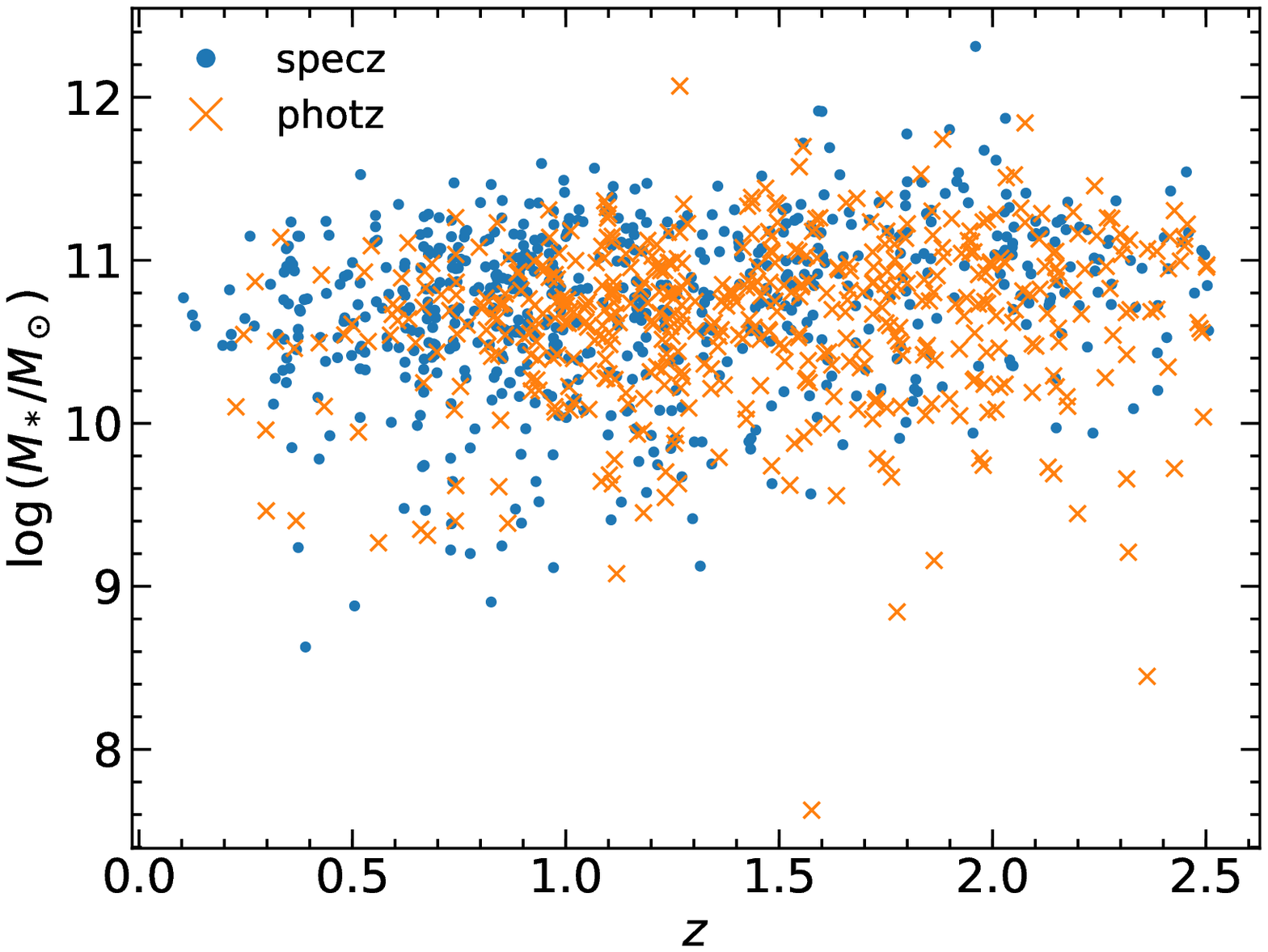}
    }
    \caption{Distribution of 2-10 keV luminosity (left) and host galaxy stellar
    mass (right) as a function of redshift for our sample of $1130$ AGNs. Blue
    (orange) points show 632 (498) AGN with known spectroscopic (photometric)
    redshifts.}
    \label{fig:distribution_z_lx_mstar}
\end{figure*}

In this paper we use the catalog presented in \citet{bongiorno12}, and we focus
on $1130$ AGN in the redshift range $0.1 < z < 2.5$, with mean $z \sim 1.2$. The
redshifts are either spectroscopic (632) or high quality photometric (498) ones.
The 2-10 keV luminosity $L_X$ spans $\log (L_X / \,\mathrm{erg}
\,\mathrm{s}^{-1}) = 42.3-45.5$ with a mean $\log (L_X / \,\mathrm{erg}
\,\mathrm{s}^{-1}) = 43.7$. The typical host galaxy of our AGN is a red and
massive galaxy with mean $\log \left( M_* / \,\mathrm{M}_\sun \right) = 10.7$.
However, the host galaxies also span a wide range of stellar masses with $\log
\left( M_* / \,\mathrm{M}_\sun \right) = {7.6-12.3}$. The $L_X$ and $M_*$
distributions for our sample of $1130$ \textit{XMM}-COSMOS AGN are shown in
Figure \ref{fig:distribution_z_lx_mstar}. It would be of interest to also study
the clustering as a function of host galaxy SFR or specific SFR (SFR$/M_*$) as
recently done by \cite{mountrichas19}. However, \cite{bongiorno12} conclude for
\textit{XMM}-COSMOS that while stellar masses from SED fitting are relatively
robust for both type 1 and type 2 AGNs, SFRs are more sensitive to AGN
contamination from type 1 AGN and are unreliable. Thus in order to increase
statistics in our clustering analysis, we will not consider the host galaxy SFR,
available only for type 2 AGN in \emph{XMM}-COSMOS.

The recent \textit{Chandra} COSMOS Legacy Survey \citep[CCLS;][]{civano16,
marchesi16} contains the largest sample of X-ray selected AGNs to date. However,
for CCLS AGN, host galaxy properties have only been estimated for type 2 AGNs,
while \cite{bongiorno12} provide the estimates for both type 1 and 2 AGNs.
Further, the clustering of \textit{XMM}-COSMOS AGNs is well studied
\citep{miyaji07, gilli09, allevato11, allevato12, allevato14a}, but not in terms
of host galaxy properties as in this work. For CCLS AGN, \cite{allevato16}
measured the clustering at $2.9 \leq z \leq 5.5$, and \cite{koutoulidis18} used
multiple fields including COSMOS to measure the clustering. Thus, there are no
clustering measurements for CCLS AGN at the redshift of interest ($z < 2.5$).

\subsection{AGN Subsamples}

The full AGN sample with known spectroscopic redshifts consists of $N=632$ AGNs
with mean $z = 1.19$. For AGNs with only known photometric redshifts, we take
into account the full probability distribution function $\mathrm{Pdf}(z)$. In
this picture, the total weight of an AGN is the integral over $z$. We limit
ourselves to $z < 2.5$ and the combined weighted number of AGNs with photometric
redshifts is $N = 488.64$ with weighted mean $z = 1.44$

To study the dependence on host galaxy properties, we divide our AGN sample
effectively in two bins of $M_*$ and $L_X/M_*$ which we refer to as the
\textit{low} and \textit{high} subsamples. In detail, first we bin the
distribution of host galaxy stellar mass $\log M_*$ of the sample with binsize
$0.1$ dex. Then, each bin is split individually exactly in half based on the
logarithm of the specific BH accretion rate $\log L_X/M_*$ to create the
\textit{low} and \textit{high} $L_X/M_*$ subsamples. The \textit{low} and
\textit{high} $L_X/M_*$ subsamples consist of $309$ objects each. We find the
average values for the \textit{low} (\textit{high}) $L_X/M_*$ subsamples to be
mean $\log L_X/M_* = 32.53\ (33.49)$, while the difference in mean $\log M_*$ is
$\lesssim 0.01$. We then repeat this process by binning the $\log L_X/M_*$ and
splitting in terms of $\log M_*$. The number of objects in the \textit{low} and
\textit{high} $M_*$ subsamples is $309$. The average values for the \textit{low}
(\textit{high}) $M_*$ subsamples are mean $\log M_* = 10.39\ (11.05)$ and the
difference in mean $\log (L_X/M_*)$ is $\lesssim 0.01$.

COSMOS is known to be affected by cosmic variance that influences the clustering
measurements \citep[e.g.][]{gilli09, mendez16}. This means that it is also
important to take into account how our \textit{low} and \textit{high} $M_*$ AGN
subsamples relate to the large structures in the field. To this end, as an
additional test, we associate the AGN sample with known spectroscopic redshifts
with the co-added COSMOS galaxy group catalog \citep[see][]{finoguenov07,
leauthaud10, george11}. An AGN is taken to belong to a galaxy group if the
AGN-group angular separation on the sky is $< R_{200,\mathrm{deg}}$ (radius of
the group in degrees enclosing 200 times the critical density), and the radial
comoving distance separation is $< \pi_\mathrm{max}$ (see Section
\ref{sec:methods}). We find $22$ ($17$) AGNs in our \textit{low} (\textit{high})
$M_*$ AGN subsamples with spectroscopic redshifts in galaxy groups with a total
number of $39$ AGNs.

We summarize the properties of the different AGN subsamples in Table 1, and the
$L_X$ and $M_*$ distributions are shown in Figures 2 and 3.

\begin{table*}[htbp]
    \centering
    \caption{\textit{XMM}-COSMOS AGN subsamples.
    }
    \begin{tabular}{lrrrrrrrrr}
        & $\sum_i \mathrm{pdf}_i$
        & $\mean{z}$
        & $\mean{\log \left( M_*/\mathrm{M}_\odot \right)}$
        & $\mean{\log \frac{L_X/M_*} {\,\mathrm{erg}\,\mathrm{s}^{-1}\,\mathrm{M}_\odot}}$
        & type1/type2
        & $\chi^2_\mathrm{min}$
        & dof
        & $b$
        & $\log \frac{M_\mathrm{halo}}{h^{-1}\mathrm{M}_\odot}$ \\
    Specz \\
\hline
        All
        & 632
        & \speczmeanz
        & \speczlogm
        & \speczloglxm
        & 351/281
        & \speczchisq
        & \speczdof
        & $\speczbias_{-\speczdbiaslo}^{+\speczdbiashi}$
        & $\speczmhalo_{-\speczdmhalolo}^{+\speczdmhalohi}$ \\
        Low $L_X/M_*$
        & 309
        & \speczlxmlomeanz
        & \speczlxmlologm
        & \speczlxmlologlxm
        & 88/221
        & \speczlxmlochisq
        & \speczlxmlodof
        & $\speczlxmlobias_{-\speczlxmlodbiaslo}^{+\speczlxmlodbiashi}$
        & $\speczlxmlomhalo_{-\speczlxmlodmhalolo}^{+\speczlxmlodmhalohi}$ \\
        High $L_X/M_*$
        & 309
        & \speczlxmhimeanz
        & \speczlxmhilogm
        & \speczlxmhiloglxm
        & 253/ 56
        & \speczlxmhichisq
        & \speczlxmhidof
        & $\speczlxmhibias_{-\speczlxmhidbiaslo}^{+\speczlxmhidbiashi}$
        & $\speczlxmhimhalo_{-\speczlxmhidmhalolo}^{+\speczlxmhidmhalohi}$ \\
        Low $M_*$
        & 309
        & \speczmstarlomeanz
        & \speczmstarlologm
        & \speczmstarlologlxm
        & 134/175
        & \speczmstarlochisq
        & \speczmstarlodof
        & $\speczmstarlobias_{-\speczmstarlodbiaslo}^{+\speczmstarlodbiashi}$
        & $\speczmstarlomhalo_{-\speczmstarlodmhalolo}^{+\speczmstarlodmhalohi}$ \\
        High $M_*$
        & 309
        & \speczmstarhimeanz
        & \speczmstarhilogm
        & \speczmstarhiloglxm
        & 211/ 98
        & \speczmstarhichisq
        & \speczmstarhidof
        & $\speczmstarhibias_{-\speczmstarhidbiaslo}^{+\speczmstarhidbiashi}$
        & $\speczmstarhimhalo_{-\speczmstarhidmhalolo}^{+\speczmstarhidmhalohi}$ \\
\\
Specz + Photz Pdfs \\
\hline
        All
        & 664
        & \speczphotzmeanz
        & \speczphotzlogm
        & \speczphotzloglxm
        & 372/292
        & \speczphotzchisq
        & \speczphotzdof
        & $\speczphotzbias_{-\speczphotzdbiaslo}^{+\speczphotzdbiashi}$
        & $\speczphotzmhalo_{-\speczphotzdmhalolo}^{+\speczphotzdmhalohi}$ \\
        Low $L_X/M_*$
        & 325
        & \speczphotzlxmlomeanz
        & \speczphotzlxmlologm
        & \speczphotzlxmlologlxm
        & 95/230
        & \speczphotzlxmlochisq
        & \speczphotzlxmlodof
        & $\speczphotzlxmlobias_{-\speczphotzlxmlodbiaslo}^{+\speczphotzlxmlodbiashi}$
        & $\speczphotzlxmlomhalo_{-\speczphotzlxmlodmhalolo}^{+\speczphotzlxmlodmhalohi}$ \\
        High $L_X/M_*$
        & 325
        & \speczphotzlxmhimeanz
        & \speczphotzlxmhilogm
        & \speczphotzlxmhiloglxm
        & 268/57
        & \speczphotzlxmhichisq
        & \speczphotzlxmhidof
        & $\speczphotzlxmhibias_{-\speczphotzlxmhidbiaslo}^{+\speczphotzlxmhidbiashi}$
        & $\speczphotzlxmhimhalo_{-\speczphotzlxmhidmhalolo}^{+\speczphotzlxmhidmhalohi}$ \\
        Low $M_*$
        & 323
        & \speczphotzmstarlomeanz
        & \speczphotzmstarlologm
        & \speczphotzmstarlologlxm
        & 139/184
        & \speczphotzmstarlochisq
        & \speczphotzmstarlodof
        & $\speczphotzmstarlobias_{-\speczphotzmstarlodbiaslo}^{+\speczphotzmstarlodbiashi}$
        & $\speczphotzmstarlomhalo_{-\speczphotzmstarlodmhalolo}^{+\speczphotzmstarlodmhalohi}$ \\
        High $M_*$
        & 323
        & \speczphotzmstarhimeanz
        & \speczphotzmstarhilogm
        & \speczphotzmstarhiloglxm
        & 224/99
        & \speczphotzmstarhichisq
        & \speczphotzmstarhidof
        & $\speczphotzmstarhibias_{-\speczphotzmstarhidbiaslo}^{+\speczphotzmstarhidbiashi}$
        & $\speczphotzmstarhimhalo_{-\speczphotzmstarhidmhalolo}^{+\speczphotzmstarhidmhalohi}$ \\
\\
Specz no groups \\
\hline
        Low $M_*$
        & 287
        & \specznogroupsmstarlomeanz
        & \specznogroupsmstarlologm
        & \specznogroupsmstarlologlxm
        & 130/157
        & \specznogroupsmstarlochisq
        & \specznogroupsmstarlodof
        & $\specznogroupsmstarlobias_{-\specznogroupsmstarlodbiaslo}^{+\specznogroupsmstarlodbiashi}$
        & $\specznogroupsmstarlomhalo_{-\specznogroupsmstarlodmhalolo}^{+\specznogroupsmstarlodmhalohi}$ \\
        High $M_*$
        & 292
        & \specznogroupsmstarhimeanz
        & \specznogroupsmstarhilogm
        & \specznogroupsmstarhiloglxm
        & 207/85
        & \specznogroupsmstarhichisq
        & \specznogroupsmstarhidof
        & $\specznogroupsmstarhibias_{-\specznogroupsmstarhidbiaslo}^{+\specznogroupsmstarhidbiashi}$
        & $\specznogroupsmstarhimhalo_{-\specznogroupsmstarhidmhalolo}^{+\specznogroupsmstarhidmhalohi}$ \\
    \end{tabular}
    \label{tab:results}
\end{table*}

\begin{figure*}
    \resizebox{\hsize}{!}{
        \includegraphics[width=.49\textwidth]{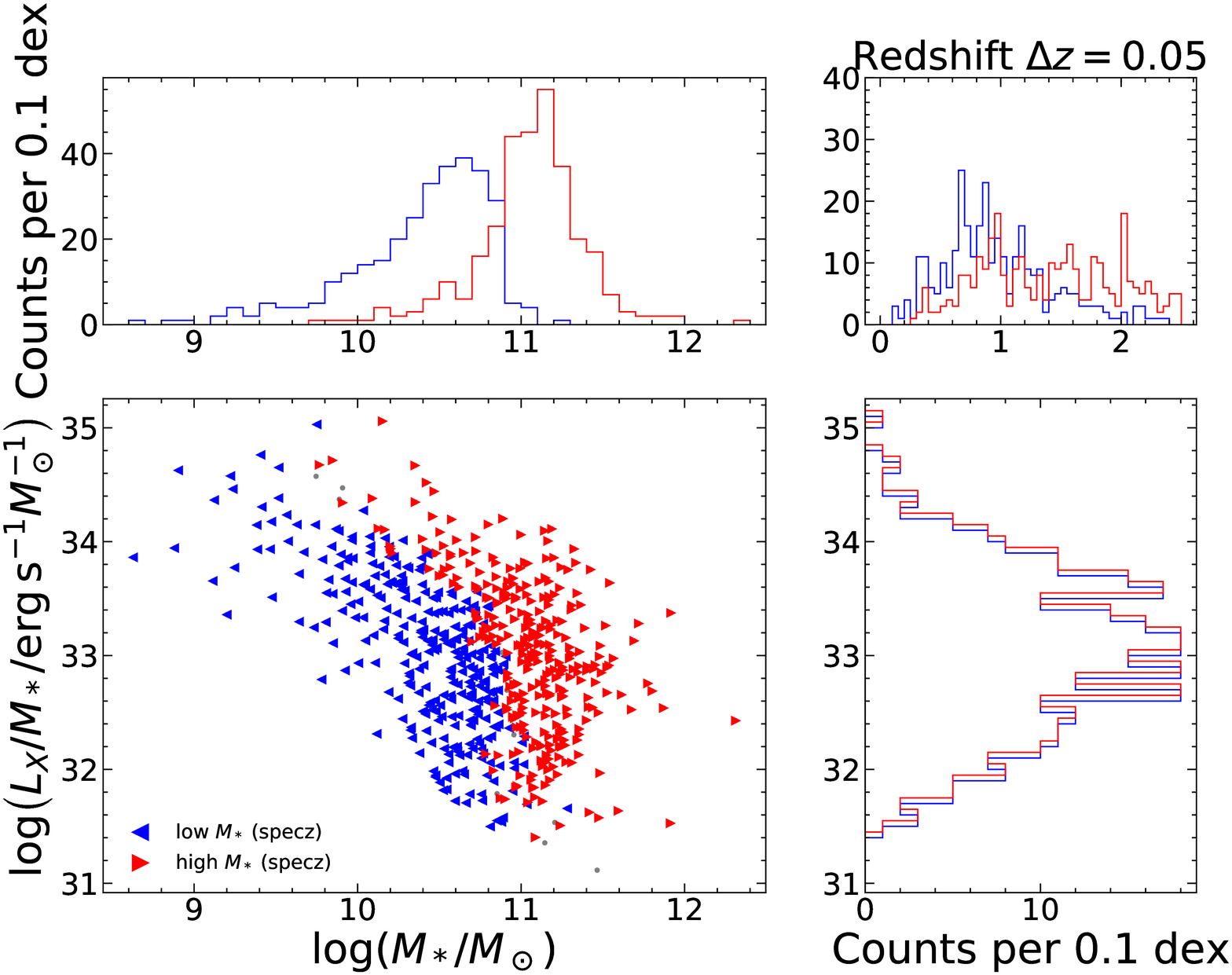}
        \includegraphics[width=.49\textwidth]{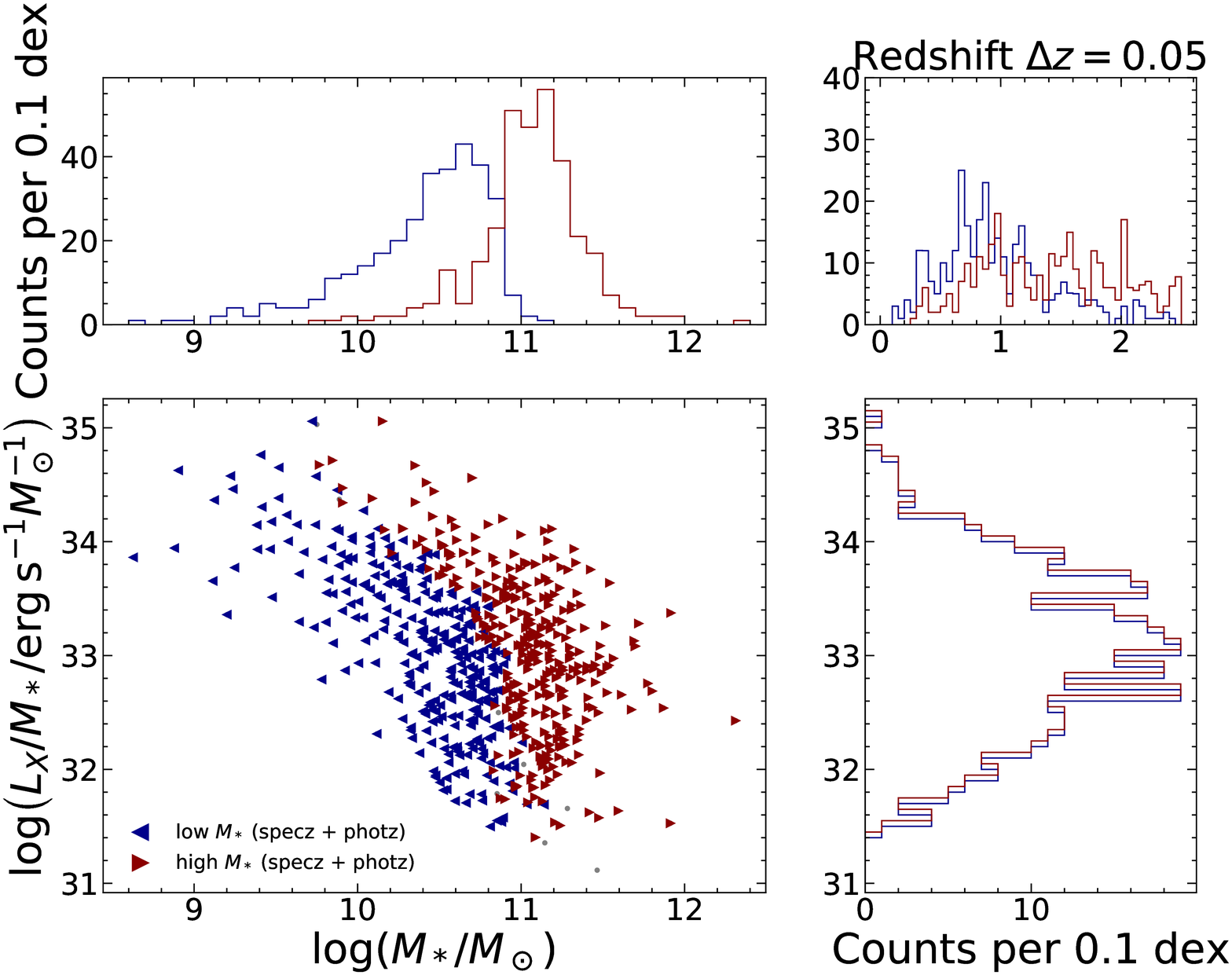}
    }
    \resizebox{\hsize}{!}{
        \includegraphics[width=.49\textwidth]{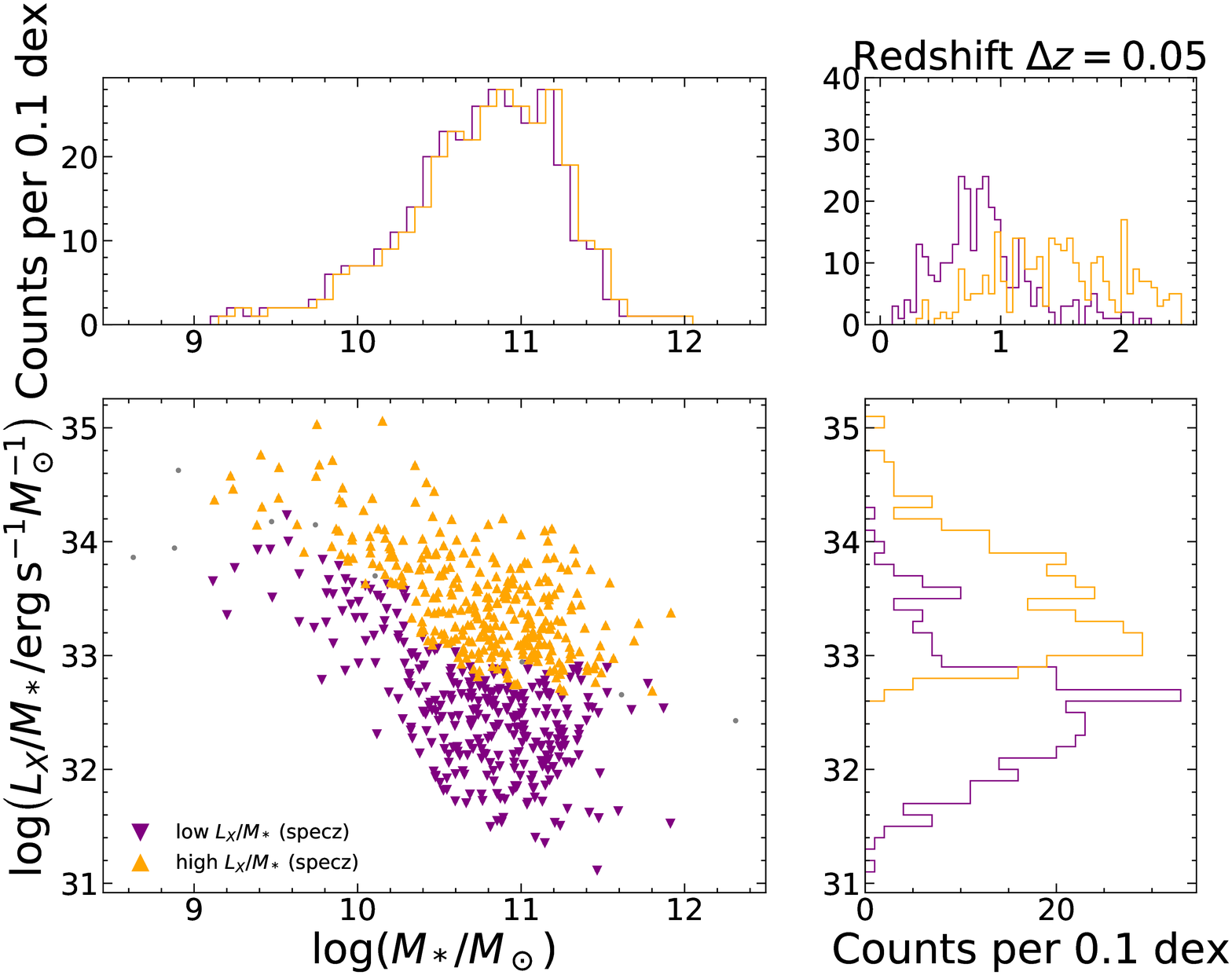}
        \includegraphics[width=.49\textwidth]{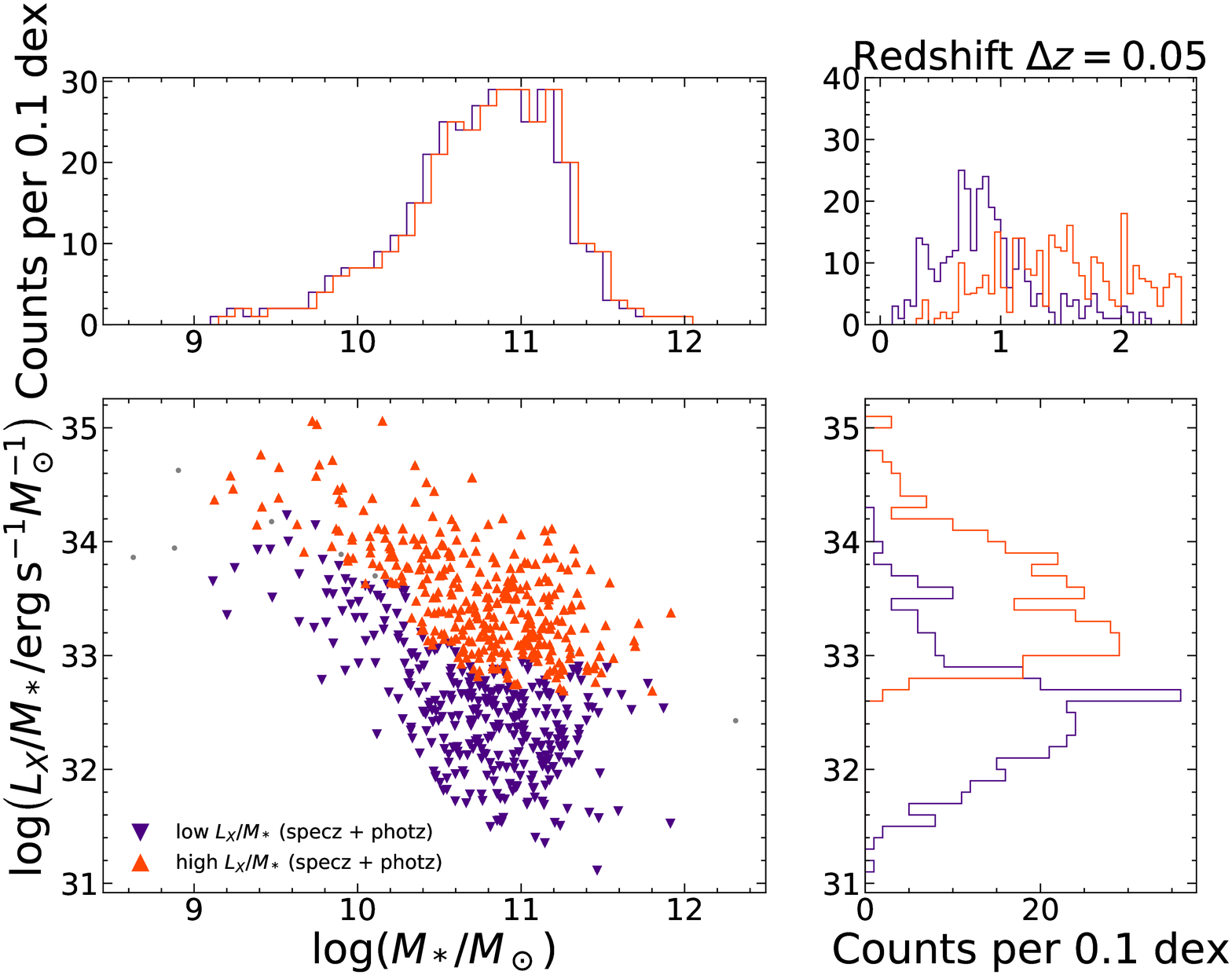}
    }
    \caption{Distribution in terms of $M_*$, $L_X/M_*$, and redshift for
    \textit{XMM}-COSMOS AGN with known spec-z (left panels) and spec-z + phot-z
    Pdfs (right panels). The \textit{low} and \textit{high} $M_*$ subsamples are
    created so that they have exactly the same specific BH accretion rate
    distribution (upper panels). A similar approach is used in terms of specific
    BH accretion rate (lower panels). For clarity, when the histograms match
    exactly, we have slightly offset the bins visually for the \textit{high}
    subsample.
    }
    \label{fig:subsamples}
\end{figure*}

\section{Methods}
\label{sec:methods}

\subsection{Two-Point Statistics}

In clustering studies, a widely used measure to quantify clustering is the
two-point correlation function $\xi(r)$ which is defined as the excess
probability above random of finding a pair of AGNs in a volume element
$\mathrm{d} V$ at physical separation $r$, so that
\begin{equation}
    \mathrm{d} P = n \left[1 + \xi(r)\right] \mathrm{d} V,
\label{eq:2pcf}
\end{equation}
where $n$ is the mean number density of AGNs. To estimate $\xi(r)$, we use the
\citet{landy_szalay93} estimator
\begin{equation}
    \xi(r) = \frac{DD' - 2DR' + RR'}{RR'},
\label{eq:lsestimator}
\end{equation}
where
\begin{align}
    DD' &= \frac{DD}{N_d(N_d - 1)/2} \\
    DR' &= \frac{DR}{N_d N_r}        \\
    RR' &= \frac{RR}{N_r(N_r - 1)/2},
\label{eq:normalized_pairs}
\end{align}
and $DD$, $DR$ and $RR$ are the number of data-data, data-random and
random-random pairs with physical separation $r$, respectively. $N_d$ and $N_r$
are the total number of sources in the data and random catalogs. This estimator
requires the creation of a random catalog to act as an unclustered distribution
of AGNs with the same selection effects in terms of RA, Dec, and redshift, as
present in the data catalog (see Section \ref{subsec:random_catalog}).

As the distances between AGN are inferred from their redshifts, the estimates
are affected by distortions due to peculiar motions of AGNs. To avoid this
effect, we express pair separations in terms of distance parallel ($\pi$) and
perpendicular ($r_p$) to the line-of-sight of the observer, defined with respect
to the mean distance to the pair. Then, the projected 2PCF, which is insensitive
to redshift space distortions, is defined as \citep{davis_peebles83}
\begin{equation}
    w_p(r_p) = 2 \int_0^\infty \xi(r_p,\pi) \mathrm{d} \pi.
\label{eq:wrp}
\end{equation}
In practice, the integration is not carried out to infinity, but to finite
value $\pi_\mathrm{max}$. The estimation of the $\pi_\mathrm{max}$ is a balance
between including all of the correlated pairs and not including noise to the
signal by uncorrelated pairs. For the estimation of the 2PCFs, we use
\texttt{CosmoBolognaLib}
\footnote{https://github.com/federicomarulli/CosmoBolognaLib}\citep{marulli16},
which is a free (as in freedom) software library for numerical cosmological
calculations.

We note that another common way to measure the clustering is to use the
cross-correlation function where positions of both an AGN sample and a complete
galaxy sample are used to decrease statistical uncertainties
\citep[e.g.][]{coil09, krumpe15, powell18, mountrichas19}. At our redshift of
interest in COSMOS, especially at $1 \lesssim z \lesssim 2.5$, it is difficult
to build a complete galaxy sample with known spectroscopic redshifts (see Sec.
\ref{subsec:generalized_estimator} for discussion on the effect of photometric
redshift in clustering measurements) to measure the clustering with, and thus we
are limited to the AGN auto-correlation function.

\subsection{Generalized Estimator}
\label{subsec:generalized_estimator}

Motivated by recent progress in utilizing photometric redshifts in AGN
clustering studies \citep{georgakakis14, allevato16}, we use the full
probability distribution function $\mathrm{Pdf}(z)$ for AGNs with no known
spectroscopic redshifts. In this approach, the classic \citet{landy_szalay93}
estimator is replaced by a generalized one, where pairs are weighted based on
$\mathrm{Pdf}(z)$ of the two objects. For the details, we refer the reader to
\citet[][Section~3]{georgakakis14}.

For the $498$ AGNs with photometric redshifts, we discretize the
$\mathrm{Pdf}(z)$ by integrating the Pdfs in terms of $z$ with an accuracy of
$\delta z = 0.01$, and normalize the Pdfs to unity. Further, we only consider
the part of the Pdf with $\mathrm{Pdf}(z) > 10^{-5}$. Using our redshift limit,
we only use the part of the Pdfs with $z < 2.5$. This means that the AGNs with
Pdfs that span over this redshift limit are cut, and for these AGNs, the Pdf
does not necessarily sum to unity i.e. $\sum_i \mathrm{Pdf}(z_i) \leq 1$.

Large uncertainties in photometric redshifts may lead to loss of not only
accuracy, but also not being able to recover the full clustering signal. This is
highlighted by the use of large values of $\pi_\mathrm{max} \gtrsim 200
\,{h}^{-1}\mathrm{Mpc}$ \citep{georgakakis14, allevato16} versus studies with
only spectroscopic redshifts with $\pi_\mathrm{max} \lesssim 100
\,{h}^{-1}\mathrm{Mpc}$ \citep[e.g.][]{coil09, allevato11, mountrichas16}.
Therefore, we select only Pdfs based on the following quality criteria: the
comoving distance separation between the $z_\mathrm{min}$ and $z_\mathrm{max}$
may not exceed a critical value of $\Delta d = 100 \,{h}^{-1}\mathrm{Mpc}$. We
define $z_\mathrm{min}$ and $z_\mathrm{max}$ separately for each AGN so that
$\mathrm{Pdf}(z) < 10^{-5}$ for $z < z_\mathrm{min}$ and $z > z_\mathrm{max}$.

In detail, from the total of 498 AGN with photometric redshifts, 32 AGN pass the
quality criterion and are included in the subsample including spectroscopic and
photometric redshifts. In terms of our $L_X/M_*$ ($M_*$) AGN subsamples, a total
of 32 (28) AGN with photometric redshifts are kept and divided equally between
the \emph{low} and \emph{high} subsamples in both cases. The number of AGN in
each of our subsamples including photometric redshifts are shown in Table
\ref{tab:results}.

This quality cut is suggested by the fact that including all phot-z Pdfs will
lead to large uncertainties in the measured clustering signal for all the AGN
subsamples. The investigation of quality criteria for studies including phot-z
Pdfs is beyond the scope of this work. However, given the importance of
photometric redshifts in future large surveys such as eROSITA, we will explore
clustering photz Pdfs in a future study (Viitanen et al., in prep.).

\subsection{Halo model}

In the halo model \citep[e.g.][]{cooray_sheth02}, the AGN clustering signal is
the sum of the 1-halo and 2-halo terms, which arise from the clustering of AGN
that occupy the same halo, and two distinct halos, respectively. On large scales
($r_p \gtrsim 1 \,{h}^{-1}\mathrm{Mpc}$), the 2-halo term is the dominant term,
and the AGN projected 2PCF may be related to the underlying DM projected 2PCF
$w_{\mathrm{DM}}^\mathrm{2-halo}$ via the linear bias $b$
\begin{equation}
    w_p^\mathrm{2-halo}(r_p) = b^2 w_{\mathrm{DM}}^\mathrm{2-halo}(r_p),
    \label{eq:bias}
\end{equation}
where $w_{\mathrm{DM}}^\mathrm{2-halo}$ is estimated at the mean redshift of the
corresponding AGN subsample and integrated to the same value of
$\pi_\mathrm{max}$. The DM projected 2PCF is related to the DM one-dimensional
2PCF $\xi_\mathrm{DM}^\mathrm{2-halo}$
\begin{equation}
    w_{\mathrm{DM}}^\mathrm{2-halo}(r_p)
    = 2 \int_{r_p}^\infty
      \frac{\xi_\mathrm{DM}^\mathrm{2-halo}(r) r \mathrm{d} r}
           {\sqrt{r^2 - r_p^2}},
\end{equation}
where $\xi_\mathrm{DM}^\mathrm{2-halo}(r)$ is in turn estimated using the linear
power spectrum $P^{\mathrm{2-halo}}(k)$:
\begin{equation}
    \xi_\mathrm{DM}^\mathrm{2-halo}(r)
    =
    \frac{1}{2\pi^2} \int P^\mathrm{2-halo}(k) k^2 \left[ \frac{\sin kr}{kr}
    \right] \mathrm{d} k.
\end{equation}
We base our estimation of the linear power spectrum on \citet{eisenstein_hu99},
which is also implemented in \texttt{CosmoBolognaLib}.

The 1-halo term ($r_p \lesssim 1 \,{h}^{-1}\,\mathrm{Mpc}$) also contains
important information on the AGN halo occupation and could be contributing
towards the clustering signal up to scales $r_p \sim 3
\,{h}^{-1}\,\mathrm{Mpc}$. However, due to low number counts of pairs especially
at small scales $r_p \lesssim 3 \,{h}^{-1}\,\mathrm{Mpc}$ in our
\textit{XMM}-COSMOS subsamples (see Fig. \ref{fig:wrp_subsamples}), we are not
able to constrain the AGN 1-halo term and excluding the 1-halo term from the
modeling does not affect our results significantly at large scales.

\subsection{Random catalog and error estimation}
\label{subsec:random_catalog}

\begin{figure*}[htbp]
    \resizebox{\hsize}{!}{
        \includegraphics[width=.24\textwidth]{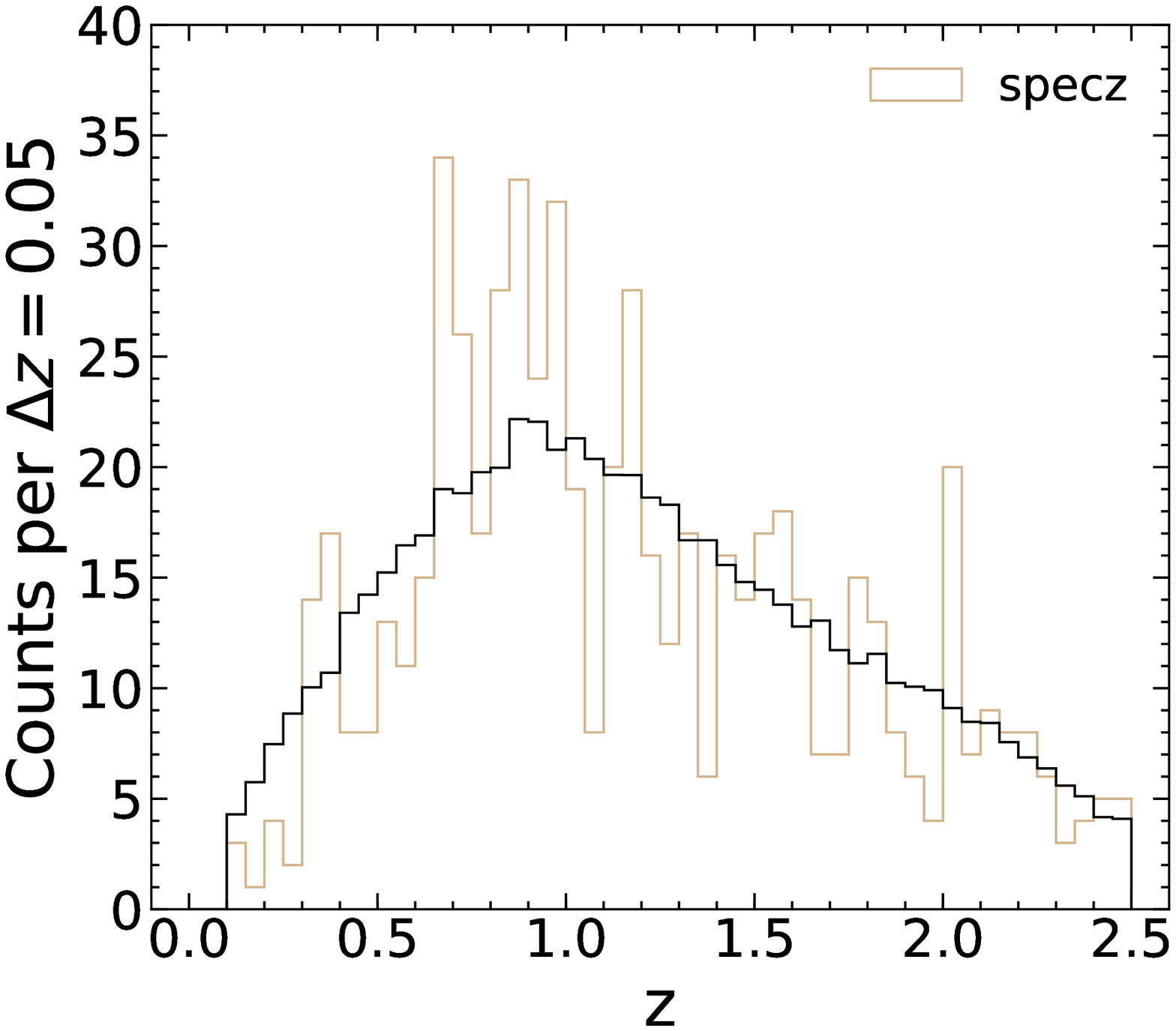}
        \includegraphics[width=.24\textwidth]{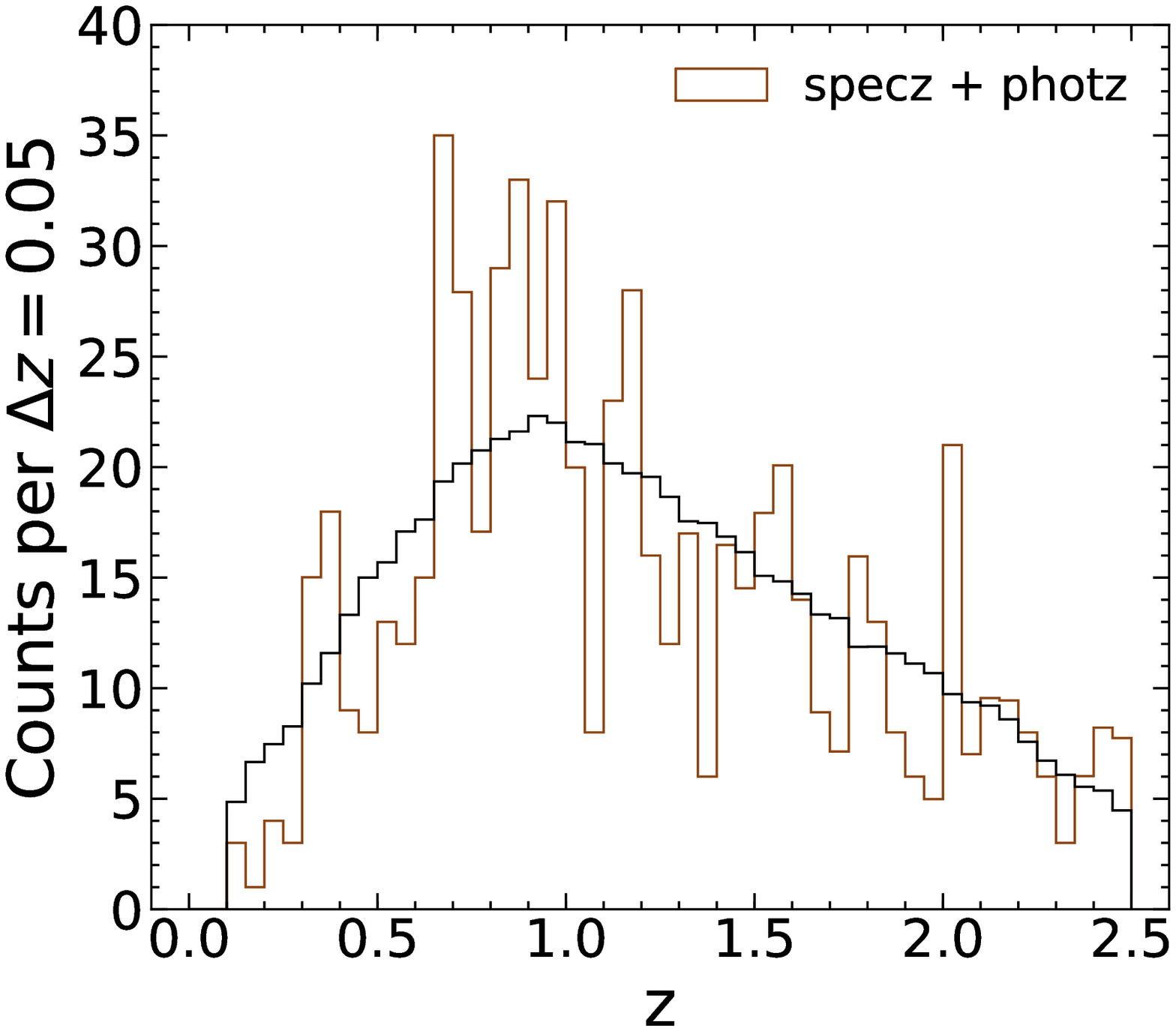}
    }
    \resizebox{\hsize}{!}{
        \includegraphics[width=.24\textwidth]{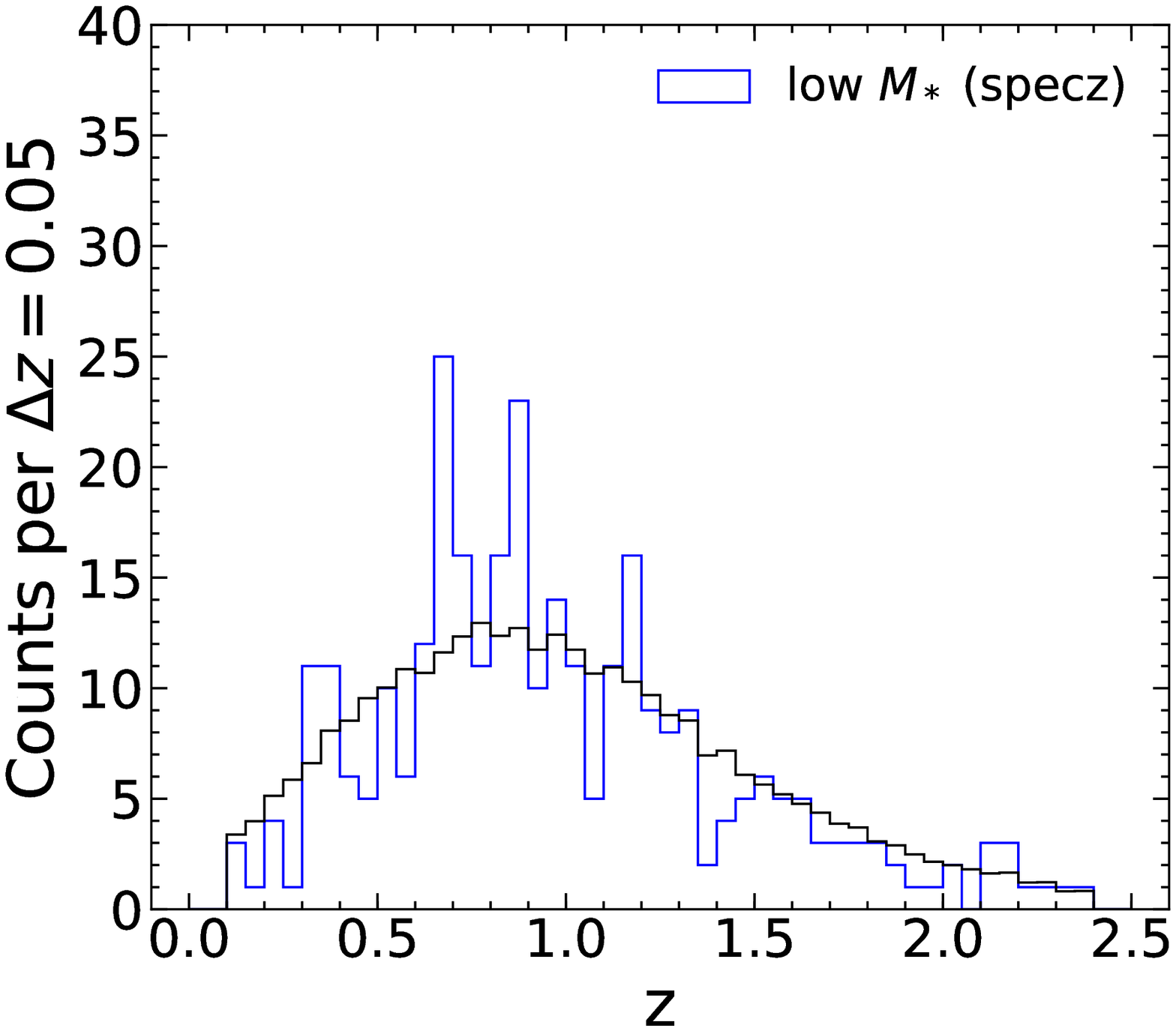}
        \includegraphics[width=.24\textwidth]{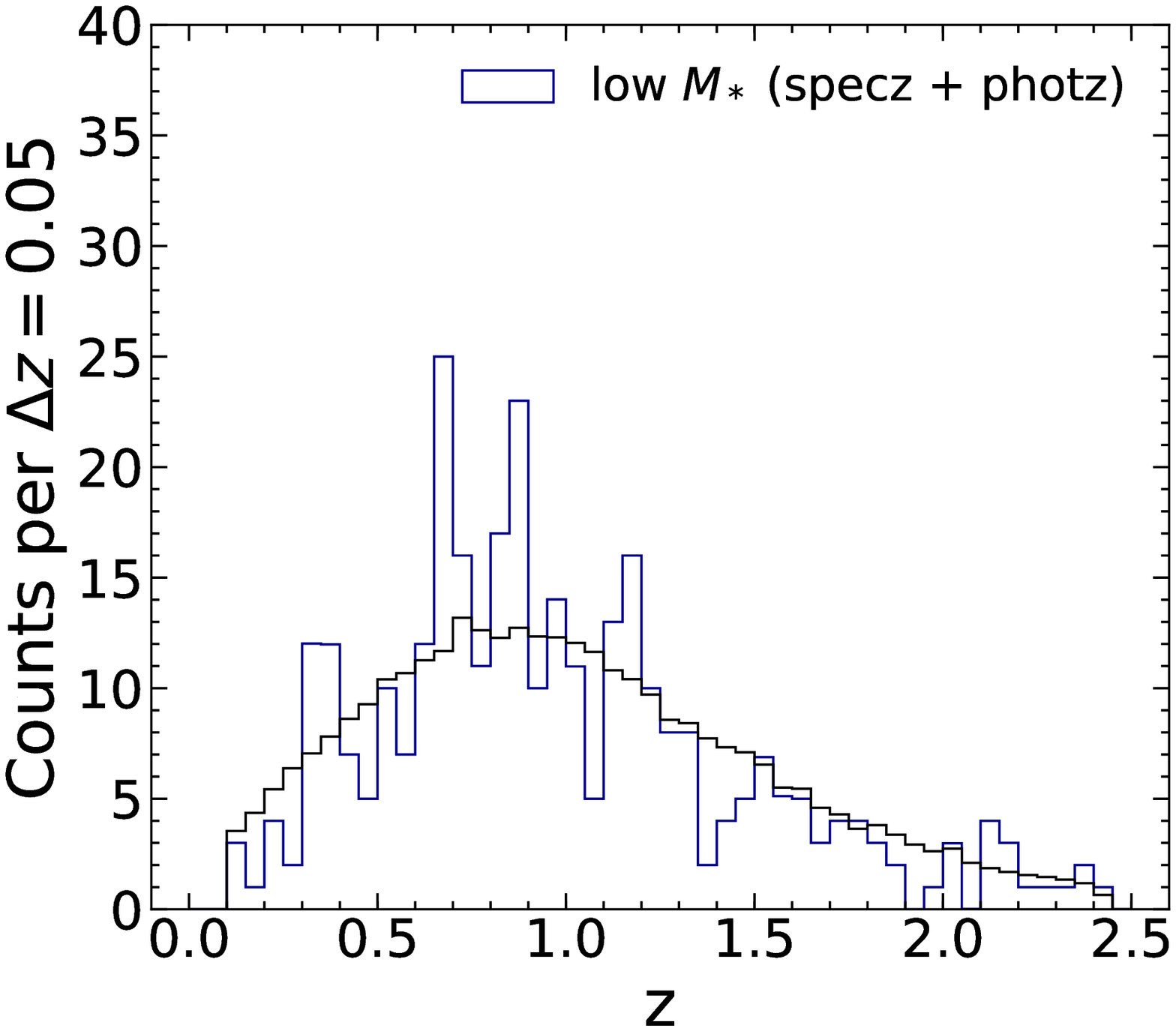}
        \includegraphics[width=.24\textwidth]{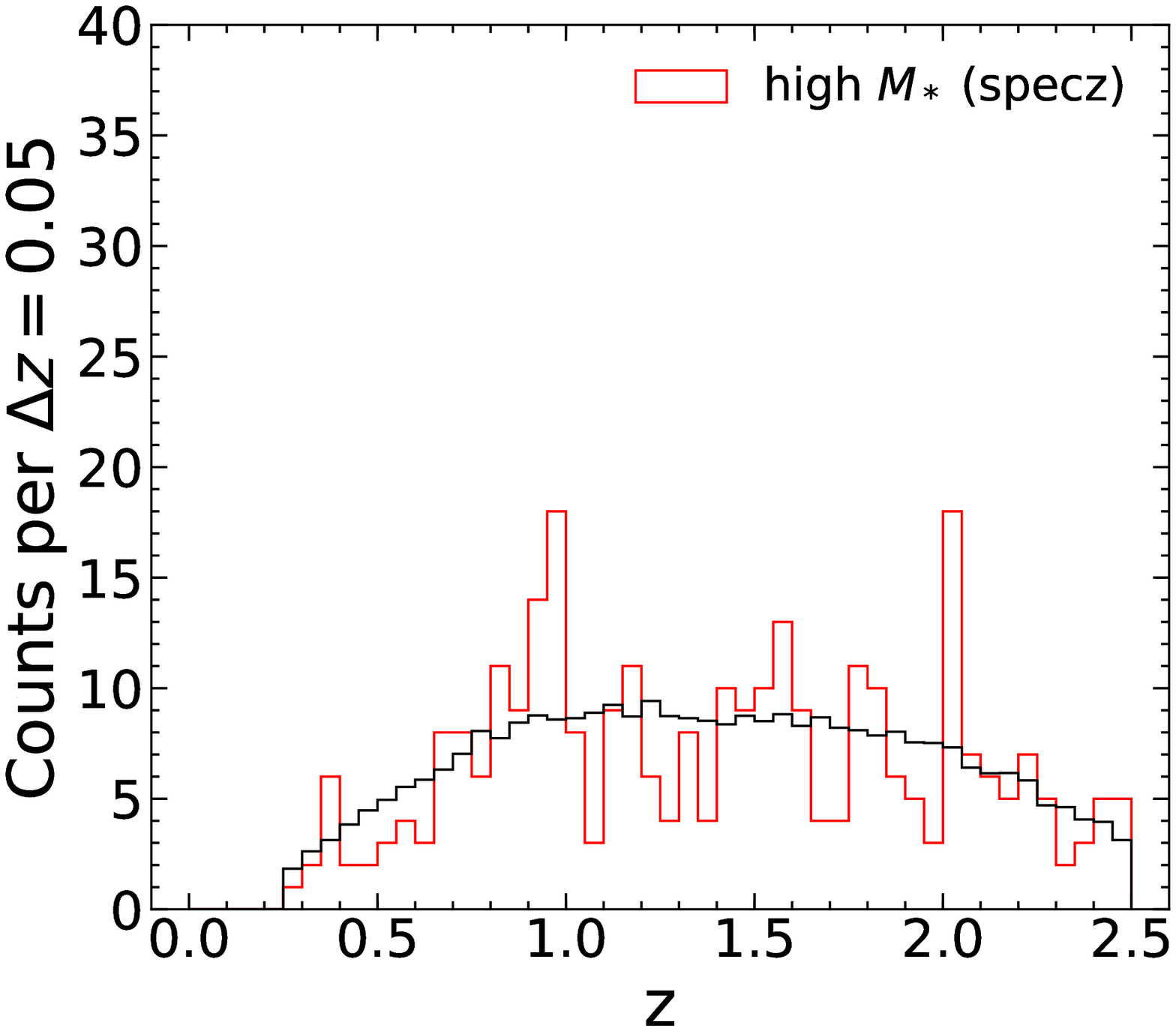}
        \includegraphics[width=.24\textwidth]{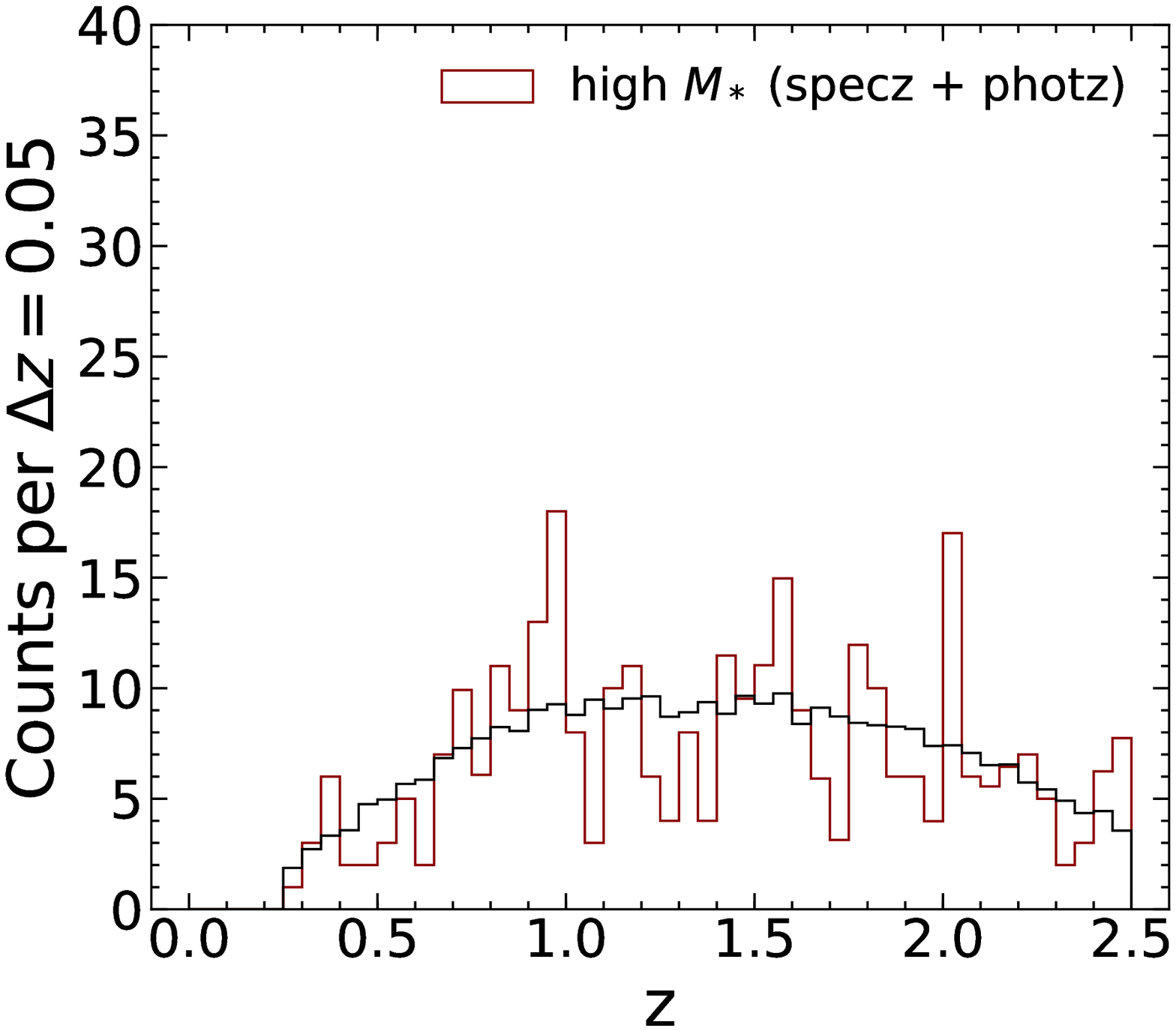}
    }
    \resizebox{\hsize}{!}{
        \includegraphics[width=.24\textwidth]{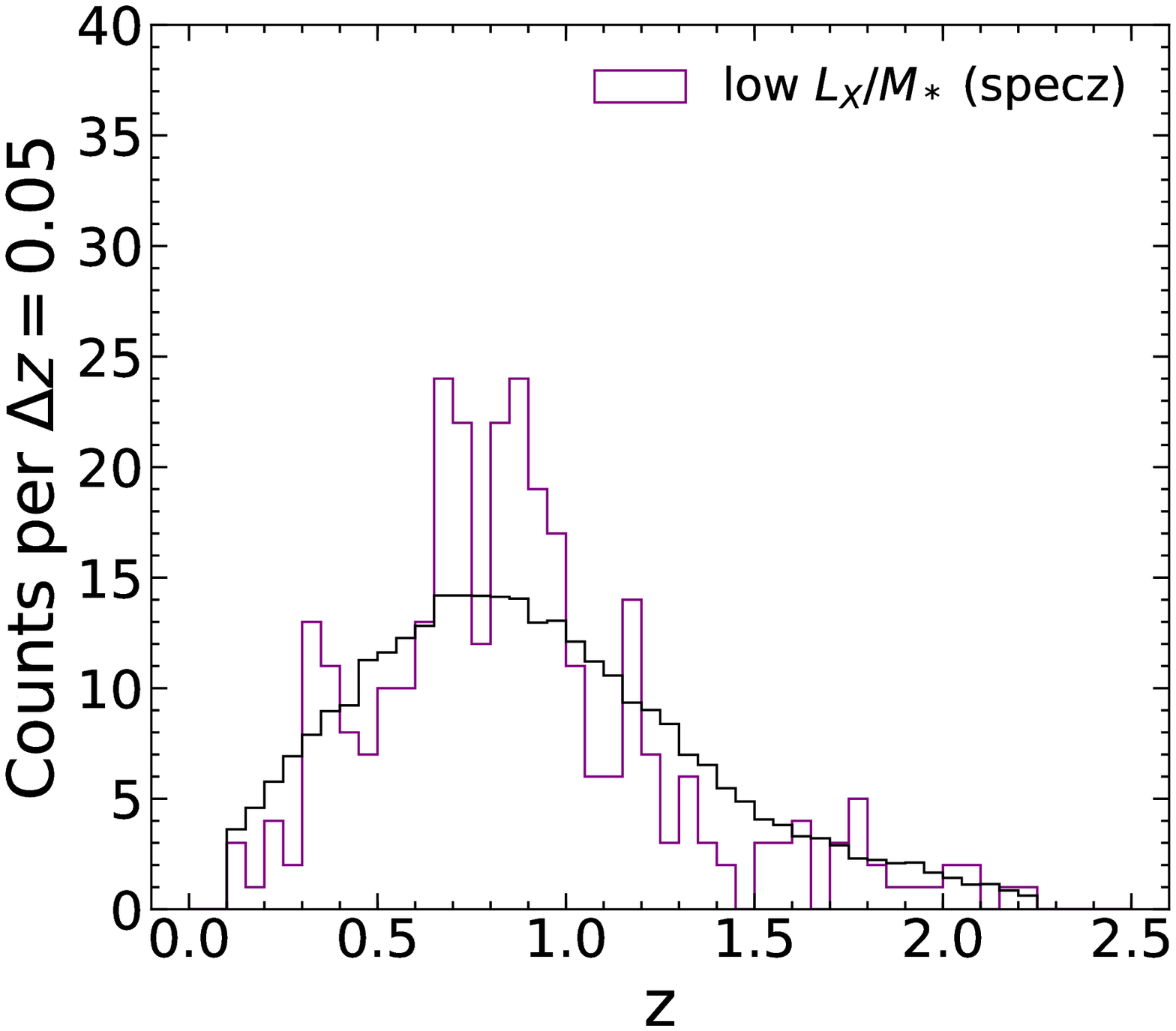}
        \includegraphics[width=.24\textwidth]{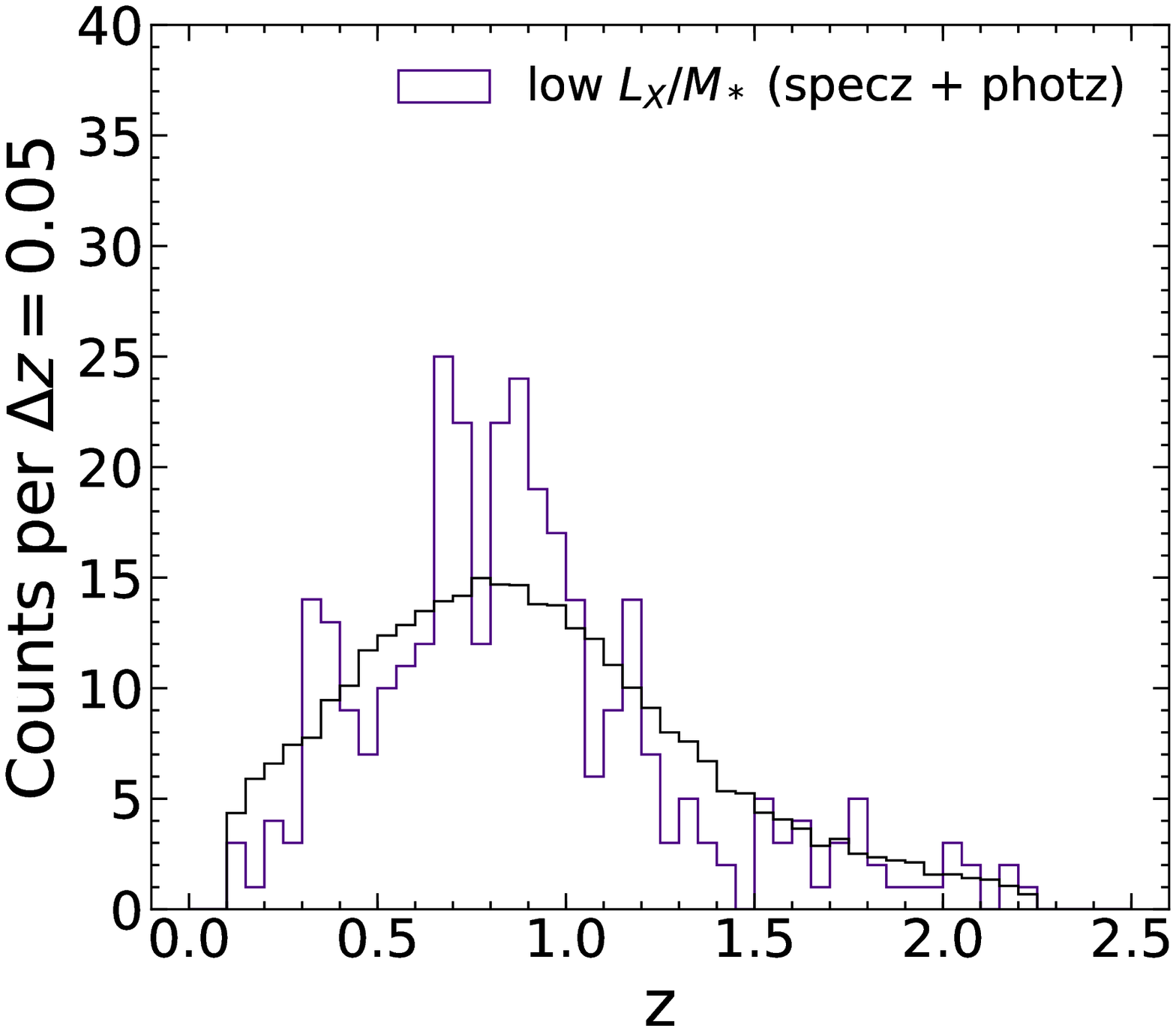}
        \includegraphics[width=.24\textwidth]{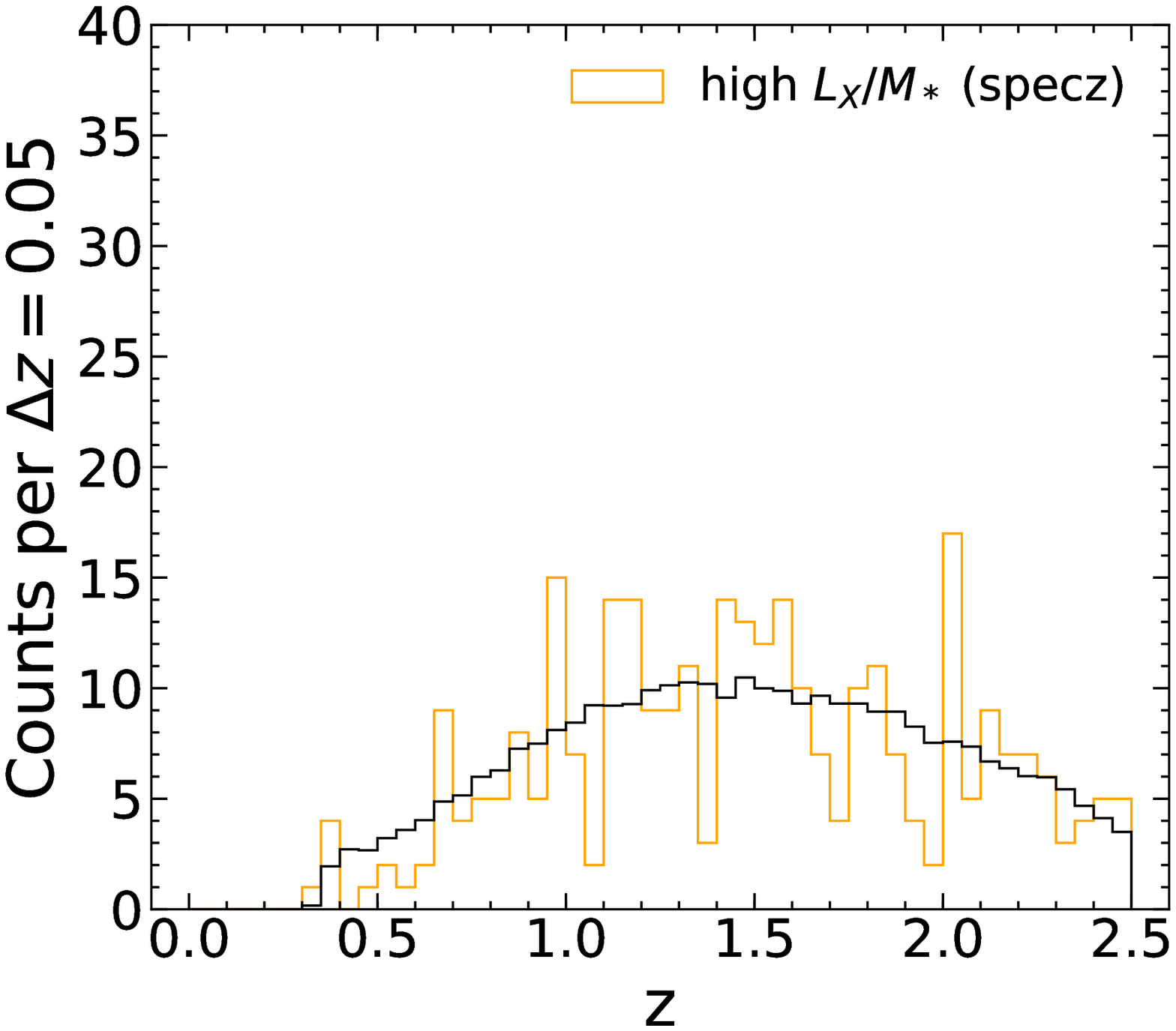}
        \includegraphics[width=.24\textwidth]{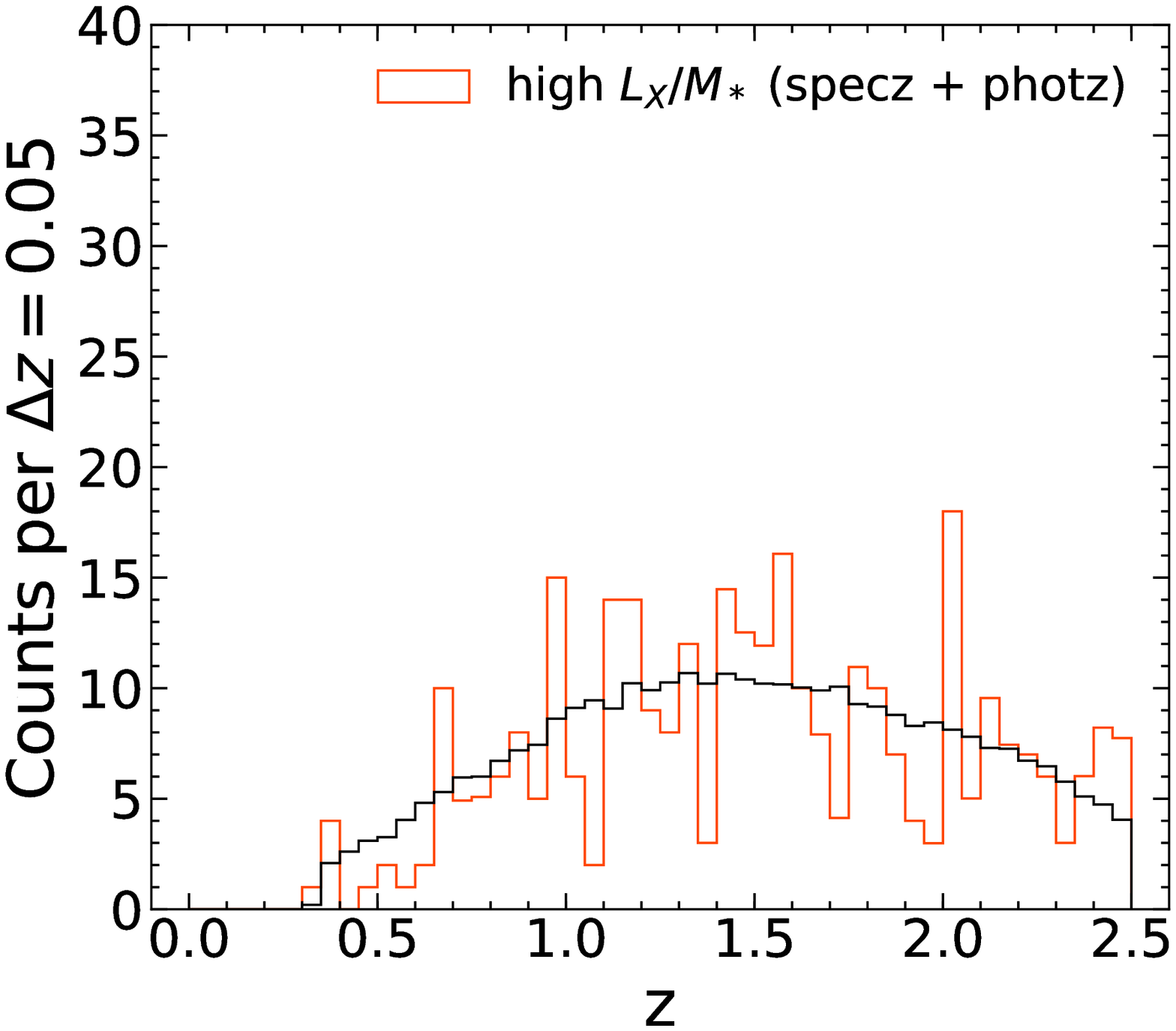}
    }
    \caption{Redshift distributions of the data and random catalogs for our AGN
    subsamples. The random redshifts are drawn from the smoothed redshift
    distribution of the data catalog using a gaussian smoothing technique with
    $\sigma_z = 0.3$.}
    \label{fig:redshift}
\end{figure*}

The random catalog consists of an unclustered set of AGNs with the same
selection effects and observational biases. To this end, we follow
\citet{miyaji07}. In detail, for each random object, we draw right ascension and
declination at random in the COSMOS field. In detail, right ascension is drawn
uniformly, while for declination we draw $\sin(\mathrm{Dec})$ uniformly. Then,
we draw a $0.5-2\,\mathrm{keV}$ flux from the data catalog, and if the drawn
flux is above the limit given by the sensitivity map \citep[][see also Figure
\ref{fig:radec}]{cappelluti09}, we keep the object. Otherwise we discard it.
Each kept random object is given a redshift drawn from the smoothed redshift
distribution of the data catalog with gaussian smoothing using $\sigma_z = 0.3$.
For each of the data catalogs, we create a random catalog with $N_r = 100 N_d$.
We show the redshift distribution of the data and random catalogs for our AGN
subsamples in Figure \ref{fig:redshift}.

Poissonian errors are readily assigned to the projected 2PCF, but are known to
underestimate the errors. For this reason we adopt a Bootstrap resampling
technique by dividing the \textit{XMM}-COSMOS survey into $N_\mathrm{region}=18$
subregions ($3\times3\times2$ for RA, Dec, and comoving distance, respectively)
of roughly equal comoving volumes. We resample the regions
$N_\mathrm{rs}=100$ times. In each of the resamples, the regions are
assigned different weights based on the number of times they are selected
\citep[][]{norberg09}. The elements of the covariance matrix $C$ are then
defined as
\begin{equation}
    C_{ij}
    =
    \frac{1}{N_\mathrm{rs}} \sum_{k=1}^{N_\mathrm{rs}}
    \left[ w_{p,k}(r_{p,i}) - \mean{w_p}(r_{p,i}) \right]
    \left[ w_{p,k}(r_{p,j}) - \mean{w_p}(r_{p,j}) \right],
    \label{eq:covariance}
\end{equation}
where $i$ and $j$ refer to the $i$th and $j$th $r_p$ bins and the bar denotes
the mean over $N_\mathrm{region}$ resamples. The $1\sigma$ error for
$w_p(r_{p,i})$ is the square root of the corresponding diagonal element i.e.
$\sqrt{C_{ii}}$.

\section{Results}

\begin{figure*}[htbp]
    \resizebox{\hsize}{!}{
        \includegraphics[width=.33\textwidth]{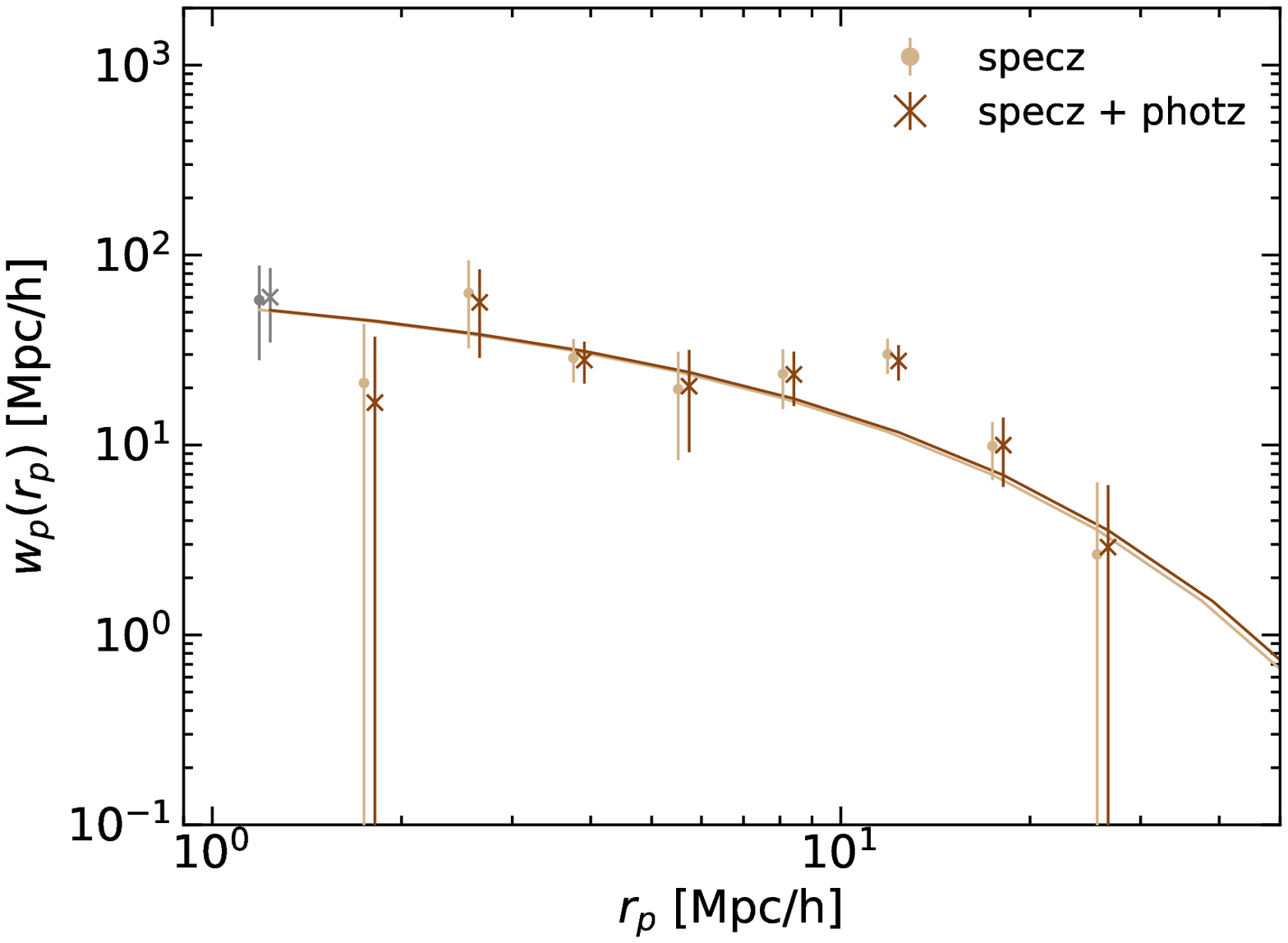}
    }
    \resizebox{\hsize}{!}{
        \includegraphics[width=.33\textwidth]{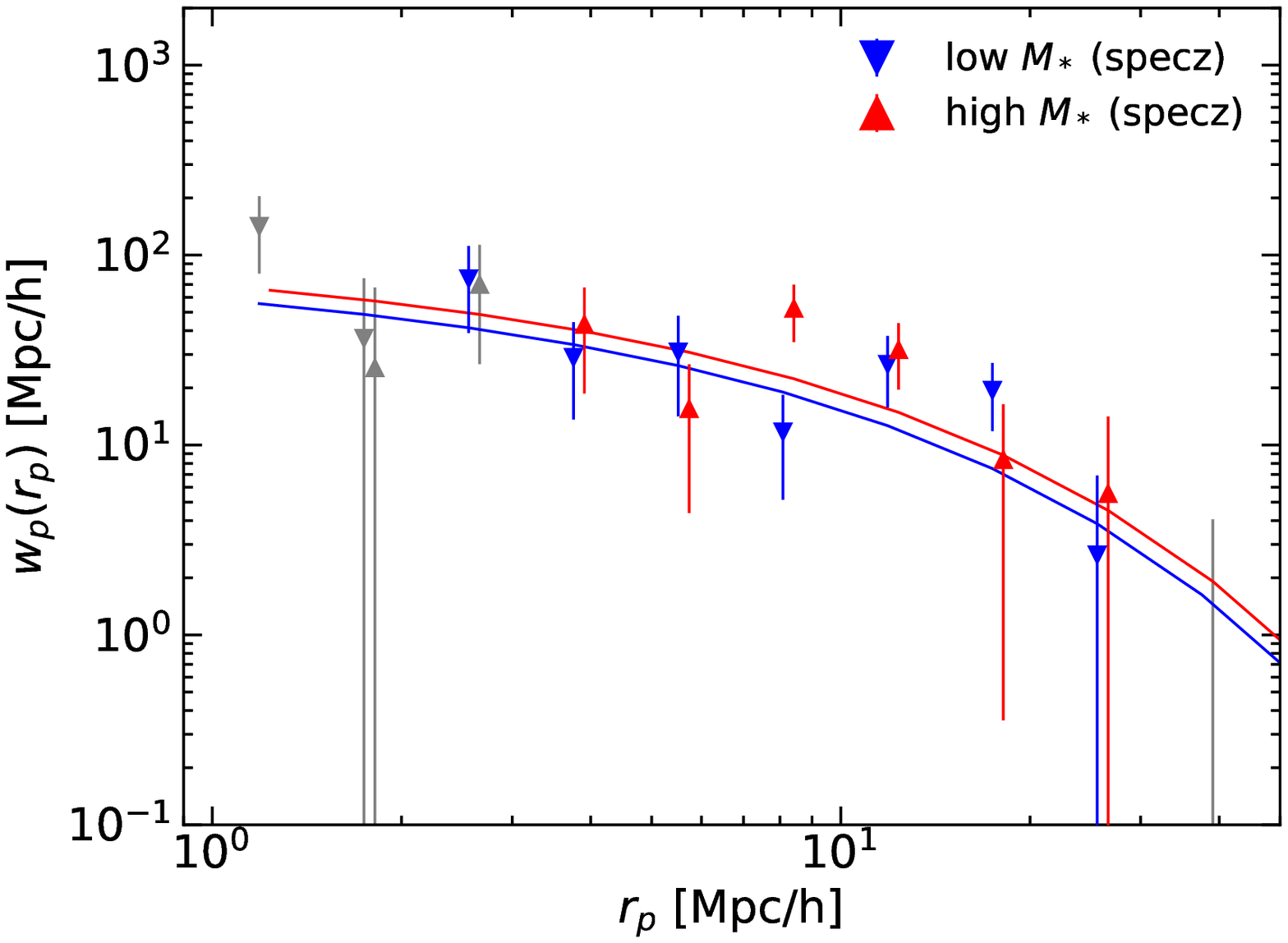}
        \includegraphics[width=.33\textwidth]{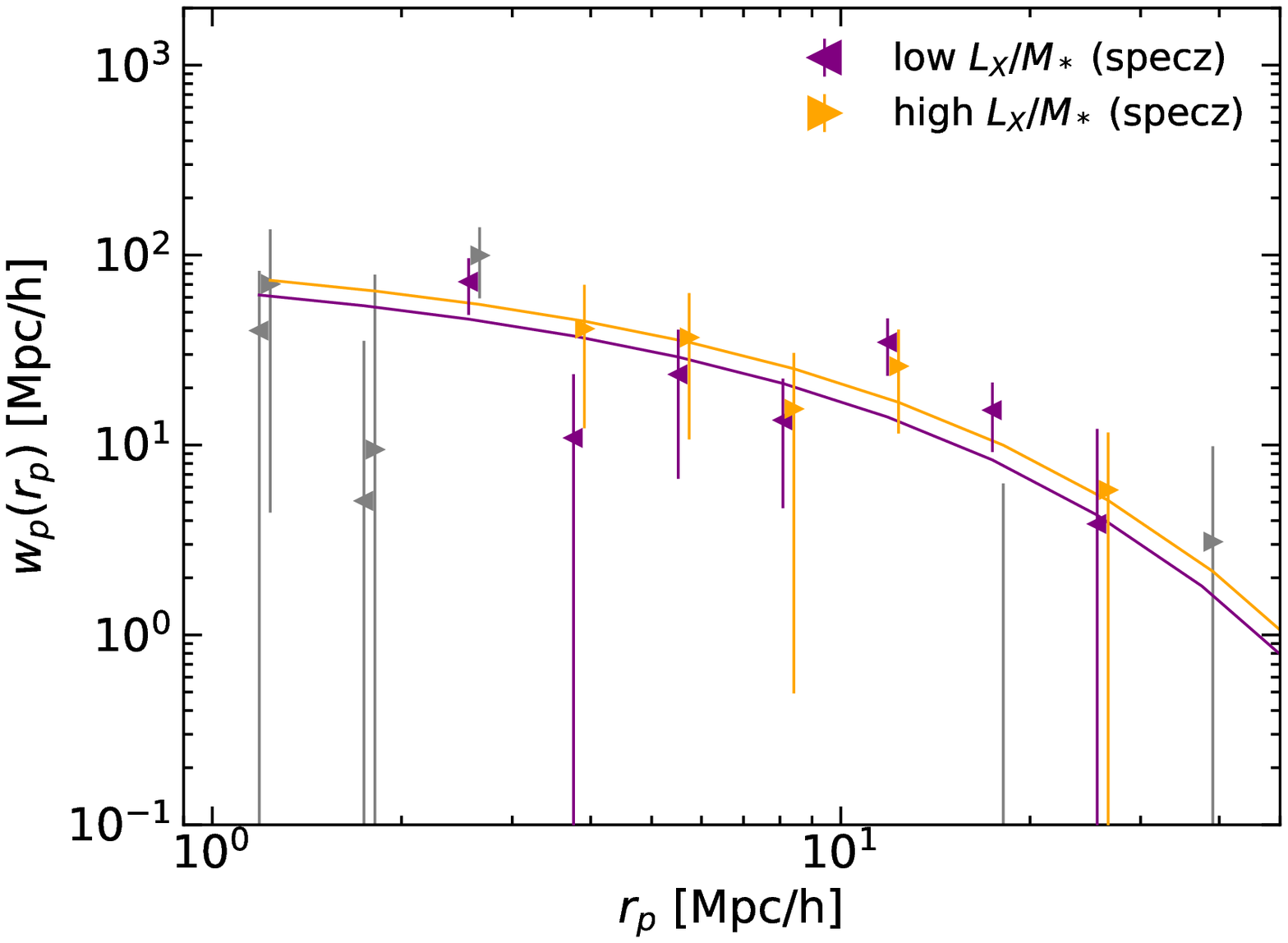}
    }
    \caption{The measured projected 2PCF for the full sample and AGN subsamples.
    The errorbars correspond to $1\sigma$ estimated via the bootstrap method.
    The solid lines show the squared best-fit bias times the projected
    DM correlation function estimated at the mean redshift of the particular
    sample. The grey datapoints are not used in the fit due to low number of
    pairs. The excess correlation at $r_p \sim 15 \,{h}^{-1}\mathrm{Mpc}$ is
    likely driven by large structure in the COSMOS field.
    }
    \label{fig:wrp_subsamples}
\end{figure*}

\begin{figure*}[htbp]
    \resizebox{\hsize}{!}{
        \includegraphics[width=.49\textwidth]{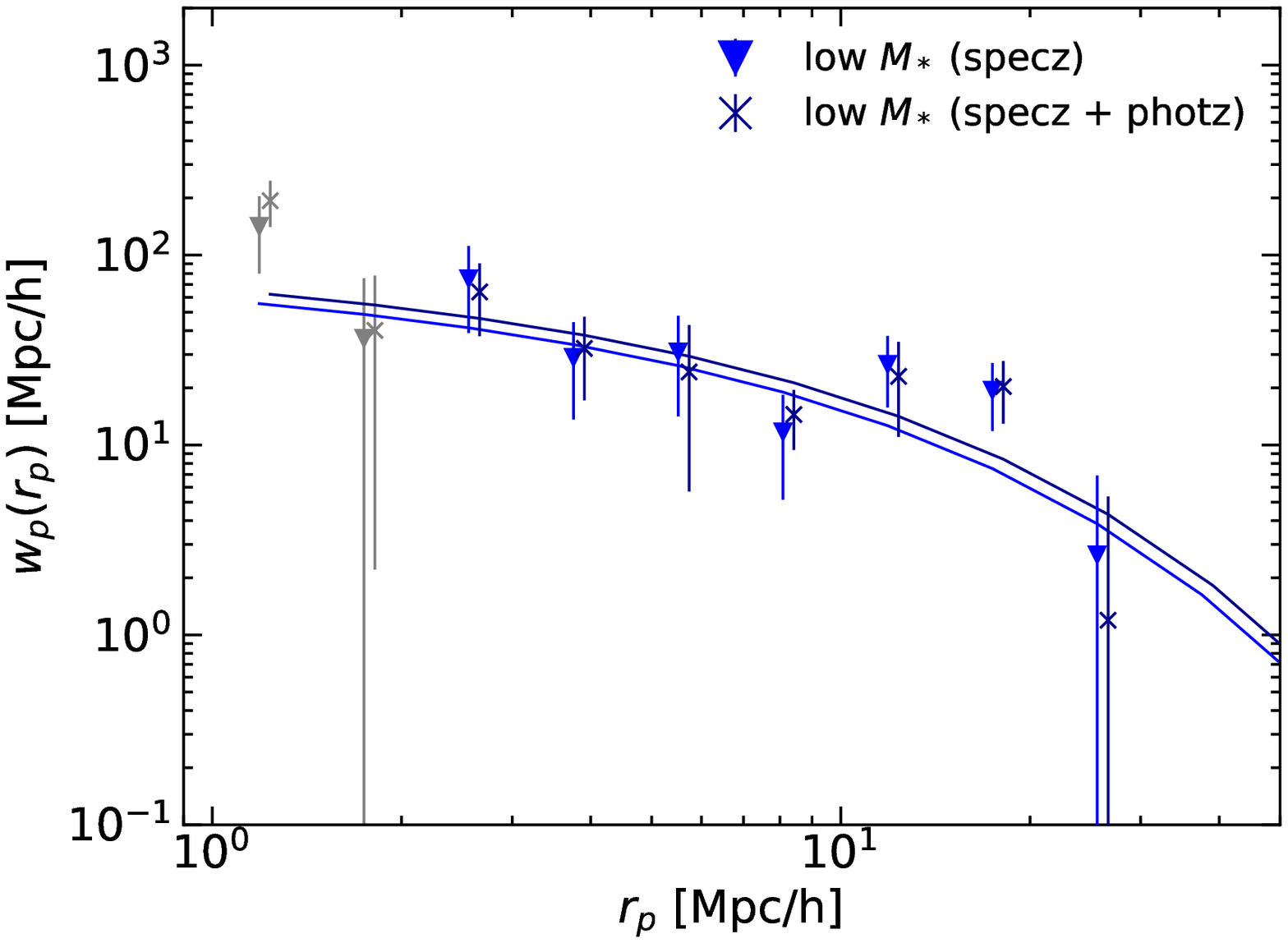}
        \includegraphics[width=.49\textwidth]{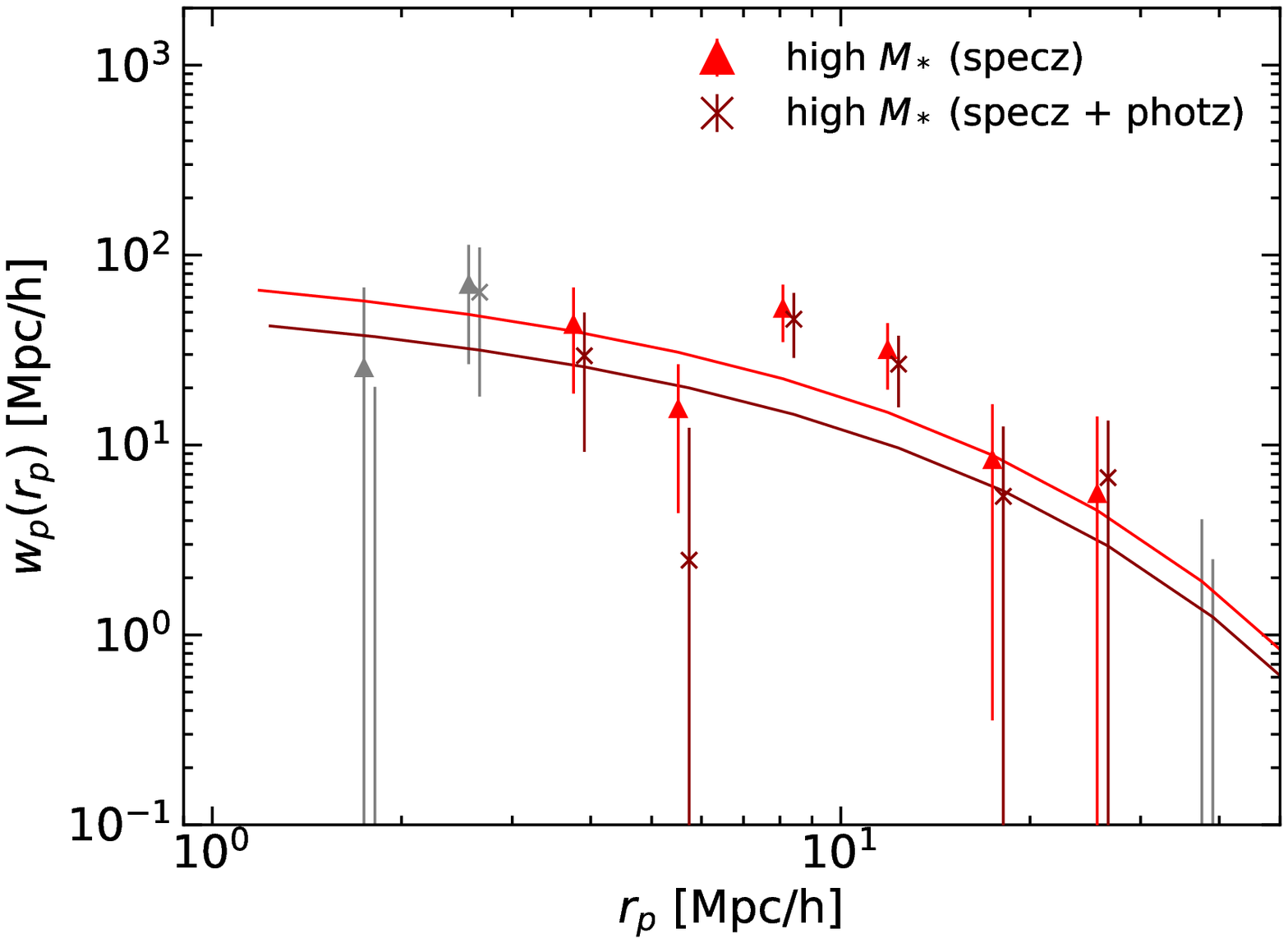}
    }
    \resizebox{\hsize}{!}{
        \includegraphics[width=.49\textwidth]{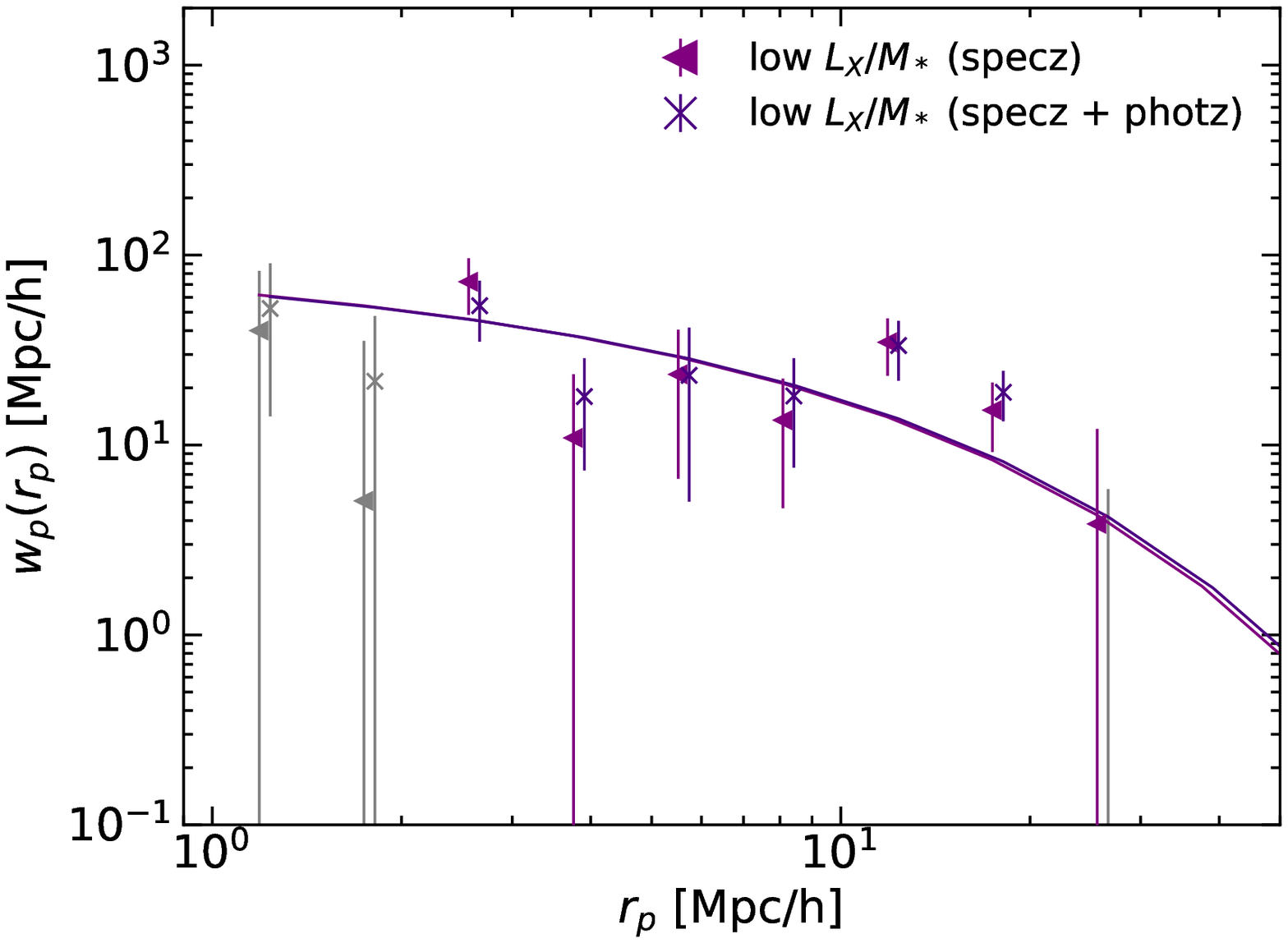}
        \includegraphics[width=.49\textwidth]{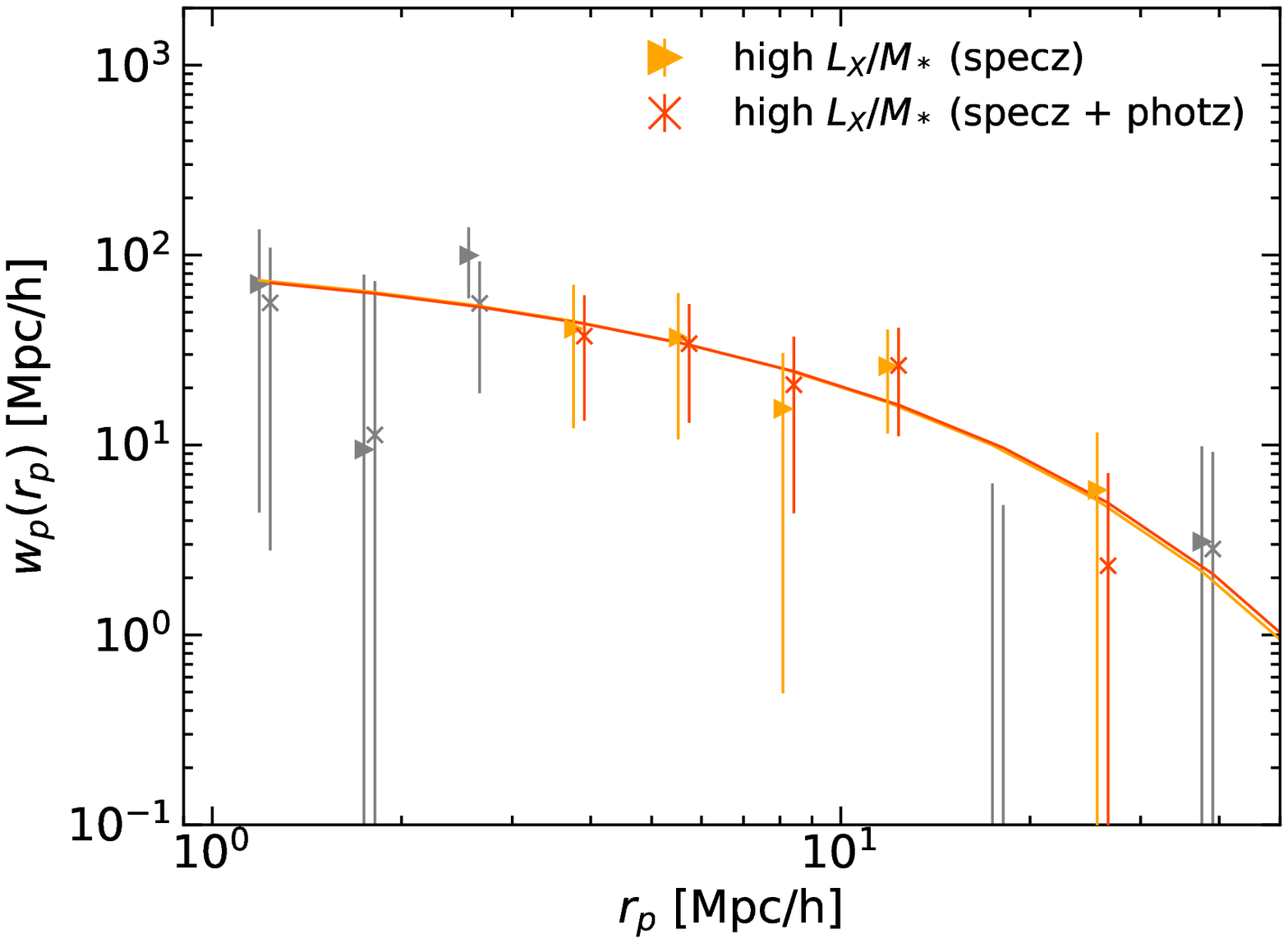}
    }
    \caption{Effect of including photometric redshifts as Pdfs in the estimation
    of the projected 2PCF (crosses). Different symbols have the same meaning as
    in Figure \ref{fig:wrp_subsamples}. The bins have been slightly offset in
    the $r_p$ direction for clarity.}
    \label{fig:wrp_photz}
\end{figure*}

For each of the AGN subsamples, we estimate the projected 2PCF $w_p(r_p)$ with
$r_p = 1.0-100 \,{h}^{-1}\mathrm{Mpc}$ using $12$ logarithmic bins. We use one
bin in the $\pi$ direction, where the upper limit of this bin is dictated by
$\pi_\mathrm{max}$. In order to set $\pi_\mathrm{max}$, we try out all the
values in the range $\pi_\mathrm{max}=20-75 \,{h}^{-1}\mathrm{Mpc}$ with an
accuracy of $\Delta \pi_\mathrm{max} = 5 \,{h}^{-1}\mathrm{Mpc}$. For the full
spectroscopic AGN sample, we found that the signal converges at
$\pi_\mathrm{max} = 40 \,{h}^{-1}\mathrm{Mpc}$, which is adopted for all the
subsamples. This value is similar to previous clustering studies involving
\textit{XMM}-COSMOS AGNs \citep{gilli09, allevato11}. The AGN projected 2PCF
$w_p(r_p)$ is then estimated using Eq. \ref{eq:wrp} and the $1\sigma$ bootstrap
errors are estimated using Eq. \ref{eq:covariance}. We show the estimated
projected 2PCF for our subsamples in Figure \ref{fig:wrp_subsamples}. Comparison
between the spectroscopic subsamples and the specz+photz subsamples are shown in
Figures \ref{fig:wrp_subsamples} (full) and \ref{fig:wrp_photz} ($M_*$ and
$L_X/M_*$ subsamples).

\begin{figure*}[htbp]
    \resizebox{\hsize}{!}{
        \includegraphics[width=.49\textwidth]{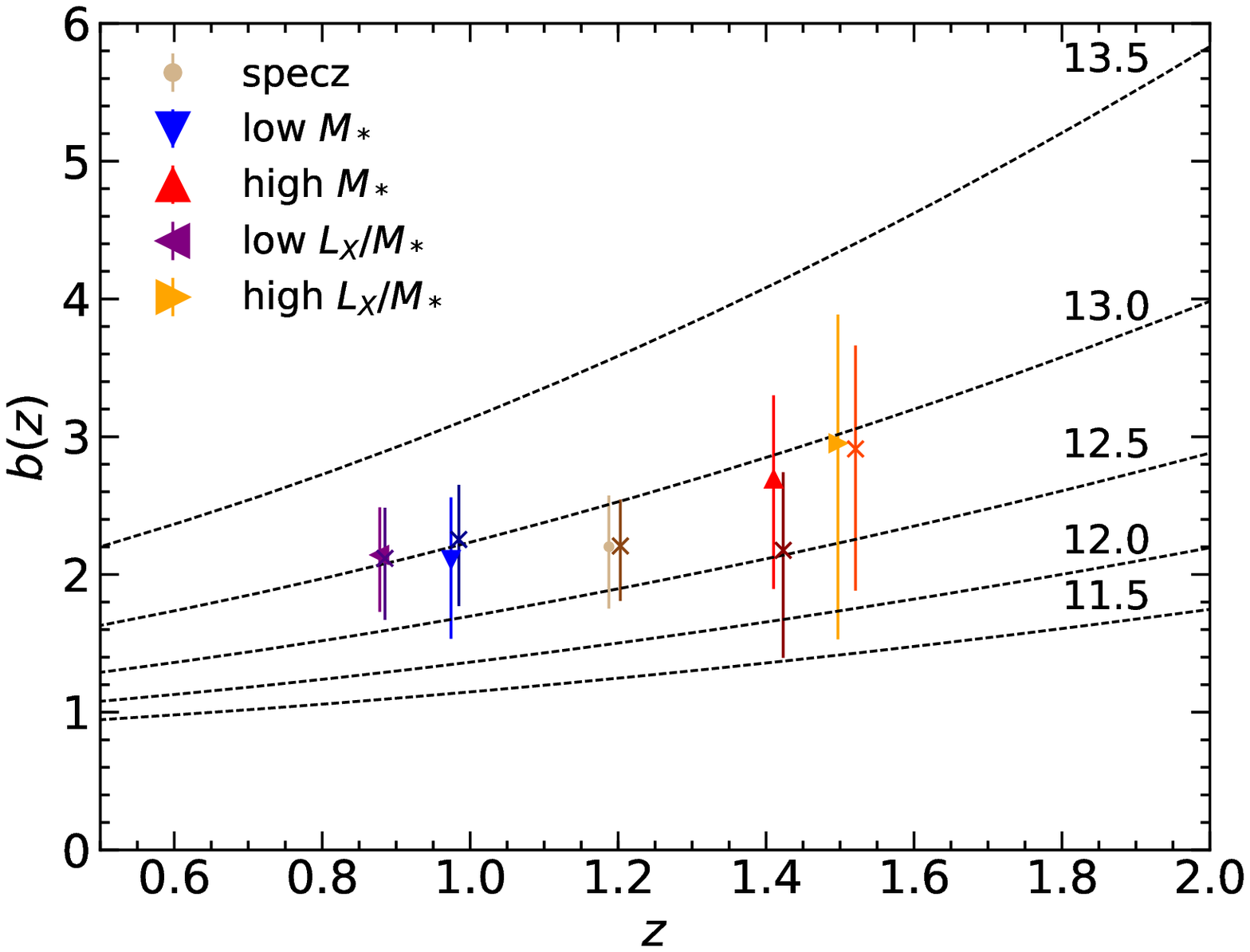}
        \includegraphics[width=.49\textwidth]{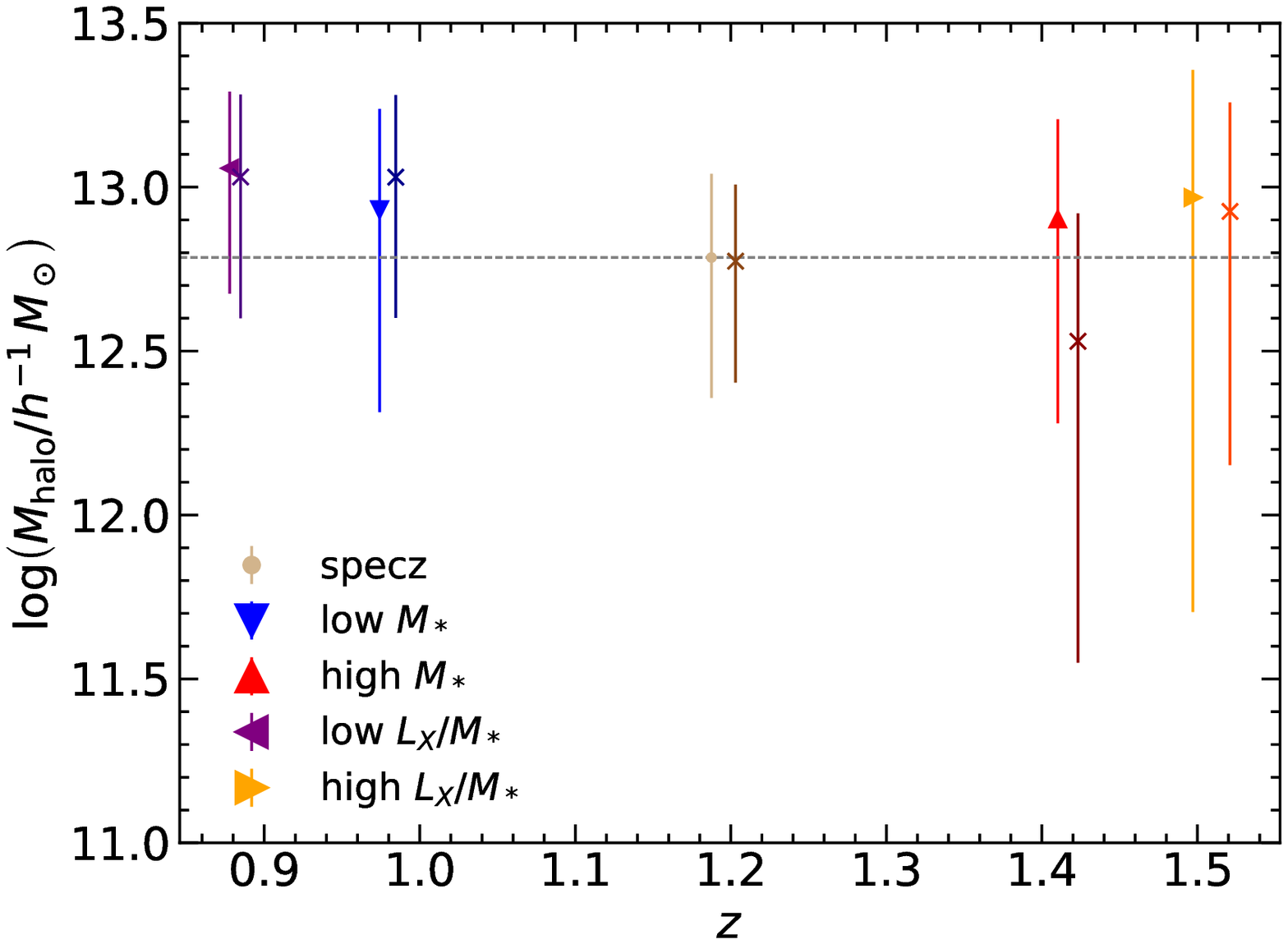}
    }
    \caption{Left: Redshift evolution of the bias for the different
    \textit{XMM}-COSMOS AGN subsamples. The grey dashed lines correspond to
    constant halo mass bias evolution $b(z, M_\mathrm{halo} = \mathrm{const})$
    for $\log M_\mathrm{halo} = 11.5,12.0,12.5,13.0,13.5$, where
    $M_\mathrm{halo}$ is given in units of $\,{h}^{-1}\,\mathrm{M}_\sun$. Right:
    Corresponding typical AGN hosting halo mass evolution with redshift. For
    visual guidance, the dashed lines show the estimated mass of the halo for
    the full spectroscopic AGN sample.}
    \label{fig:bias_halomass}
\end{figure*}

We derived the best-fit large-scale bias (Eq. \ref{eq:bias}) using $\chi^2$
minimization for $r_p = 1-30 \,{h}^{-1}\mathrm{Mpc}$. In detail, we utilize the
inverse of the full covariance matrix $C^{-1}$ and minimize $\chi^2 = \Delta^T
C^{-1} \Delta$, where $\Delta$ is a with the same number of elements as the
number of $r_p$ bins used in the fit. $\Delta$ is defined explicitly as $\Delta
= w_{p,\mathrm{AGN}}^{\mathrm{2-halo}} - b^2 w_{\mathrm{DM}}^{\mathrm{2-halo}}$.
With one free parameter, we estimate the $1 \sigma$ errors on the best-fit bias,
given by the lower and upper bounds of the region $(\chi^2 -
\chi_{\mathrm{min}}^2)/\nu \leq 1.0$, where $\nu = N - 1$ is the number of
degrees of freedom. To exclude noisy bins in the fit, we require that the number
of pairs in each bin is $>16$. The large-scale bias derived for all the
\textit{XMM}-COSMOS AGN subsamples are summarized in Table \ref{tab:results} and
shown in Figure \ref{fig:bias_halomass}.

For the full spectroscopic AGN subsample (Figure \ref{fig:wrp_subsamples}), we
find a best-fit bias of $b = \speczbias_{-\speczdbiaslo}^{+\speczdbiashi}$.
Following the bias-mass relation described in \citet{vandenbosch02} and
\citet{sheth01}, this corresponds to a typical mass of the hosting halo of
$\logMhalo = \speczmhalo_{-\speczdmhalolo}^{+\speczdmhalohi}$.
Notice that in this work we define the \emph{typical} mass explicitly as the DM
halo mass which satisfies $b = b(M_\mathrm{halo})$ \citep[e.g.][]{hickox09a,
allevato16, mountrichas19}. Albeit with large uncertainties, we find a small
$\lesssim 1\sigma$ difference in the biases of the spectroscopic AGN subsamples
split in terms of stellar mass (Figure \ref{fig:wrp_subsamples}). The biases are
$b=\speczmstarlobias_{-\speczmstarlodbiaslo}^{+\speczmstarlodbiashi}$ for the
\textit{low} stellar mass and
$b=\speczmstarhibias_{-\speczmstarhidbiaslo}^{+\speczmstarhidbiashi}$ for the
\textit{high} stellar mass. However, it is worth noting that the two subsamples
peak at different redshifts ($z \sim 1.0$ versus $z \sim 1.4$). In terms of the
typical masses of the hosting halos, we find no difference. For the $M_*$
subsamples, we find that excluding AGNs that are associated with groups has a
greater effect on the measured best-fit bias of the \textit{low} $M_*$
subsample. We measure
$b=\specznogroupsmstarlobias_{-\specznogroupsmstarlodbiaslo}^{+\specznogroupsmstarlodbiashi}$
($b=\specznogroupsmstarhibias_{-\specznogroupsmstarhidbiaslo}^{+\specznogroupsmstarhidbiashi}$)
for the \textit{low} (\textit{high}) $M_*$ AGN subsample. This lower value for
the bias could be an indication that AGNs in galaxies with lower stellar mass
are more preferably satellites in their DM halos.

Moreover, we derive an AGN bias
$b=\speczlxmlobias_{-\speczlxmlodbiaslo}^{+\speczlxmlodbiashi}$ (at $z \sim
0.9$) and $b=\speczlxmhibias_{-\speczlxmhidbiaslo}^{+\speczlxmhidbiashi}$ ($z
\sim 1.5$) for the \textit{low} and \textit{high} $L_X/M_*$ subsamples,
respectively (Figure \ref{fig:wrp_subsamples}). No significant difference is
observed in the typical masses of the hosting halos.

Similar results in terms of bias dependence on $M_*$ and $L_X/M_*$ are found
when using phot-z Pdfs in addition to any available spectroscopic redshifts. In
particular, in our full AGN subsamples, an increase of ${\sim} 5\%$ in the
weighted number of AGNs introduces no systematic error in the estimation of the
bias, but decreases the $1\sigma$ error of the bias by $(\delta b_1 - \delta
b_2) / \delta b_1 \sim 10\%$, where $\delta b_i$ is the average error derived
from the lower and upper limits of the bias (see Table \ref{tab:results}).
However, since including photometric redshifts do not change the conclusions
drawn from our clustering measurements, in the following sections we focus on
the results from the AGN subsamples with known spectroscopic redshifts.

\section{Discussion}

We have performed clustering measurements of 1130 X-ray selected AGN in
\textit{XMM}-COSMOS at $0.1 < z < 2.5$ (mean $z \sim 1.2$) in order to study AGN
clustering dependence on host galaxy stellar mass and specific BH accretion rate
$L_X/M_*$.
For our AGN subsamples we find a typical DM halo mass $\sim 10^{13}
\,{h}^{-1}\,\mathrm{M}_\sun$ that roughly correspond to group-sized
environments. This is in agreement with similar studies using X-ray selected
AGNs at similar redshifts \citep[][]{coil09, allevato11, fanidakis13,
koutoulidis13}, as well as at lower redshifts $z < 0.1$ \citep[e.g.][]{krumpe18,
powell18}. We have also investigated including photometric redshifts as Pdfs in
the analysis in addition to any available spectroscopic redshifts.

In COSMOS, \cite{leauthaud15} use weak lensing measurements on X-ray COSMOS AGN
at $z < 1$ with $\log L_X / \mathrm{erg}\,\mathrm{s}^{-1} = [41.5-43.5]$ and
$\log M_* / \mathrm{M}_\odot = [10.5-12]$. They infer that $50$ per cent of AGN
reside in halos with $\log M_\mathrm{halo} / \mathrm{M}_\odot < 12.5$ in tension
with the claim that X-ray AGN inhabit group-sized environments with masses $\sim
10^{13} \,\mathrm{M}_\odot$. However, they also underline that due to the skewed
tail in the halo mass distribution, the effective/typical halo mass derived from
clustering measurements may be markedly different from the median of the
distribution.

In fact, they found an \emph{effective} mass of $M_\mathrm{eff} \sim 10^{12.7}
\,\mathrm{M}_\odot$, which is close to the typical halo masses derived in this
work. It is worth noticing that they derived the effective halo mass from
modelling the AGN halo occupation \citep[see Eq. (4) in][]{leauthaud15}, which
may differ from the typical halo mass inferred from the 2-halo term as in this
work.

Also, they found that the effective DM halo mass of their AGN sample lies
between the median and the mean values of the DM halo mass distribution, which
are lower and higher than the effective DM halo mass, respectively. Given the
statistics in our \emph{XMM}-COSMOS AGN sample, we are not able to constrain the
median or the mean of the DM halo mass distribution. In the future this could be
done through HOD modelling, provided the 1-halo term is constrained.

Moreover, different cuts in luminosity and host galaxy mass may reflect in
different hosting DM halo mass distributions. For instance, our sample of
\emph{XMM}-COSMOS AGN spans a range of host galaxy stellar masses $\log M_*/
\mathrm{M}_\odot = [8-12]$, including also low-mass systems with masses
$< 10^{10.5} \,\mathrm{M}_\odot$ (that are likely satellite galaxies in galaxy
groups), and probes higher redshifts up to $z = 2.5$.

\subsection{Clustering in terms of specific BH accretion rate}

We divided the full sample in \textit{low} and \textit{high} specific BH
accretion rate subsamples with the same $M_*$ distributions and find no
significant clustering dependence on $L_X/M_*$, and thus Eddington ratio.
\citet{krumpe15} also found no dependence on $\lambda_\mathrm{Edd}$ for their
sample of local ($0.16 < z < 0.36$) X-ray and optically selected AGN in the
Rosat All-Sky Survey. They concluded that high accretion rates in AGN are not
necessarily linked to high density environments where galaxy interactions would
be frequent. Our result provides further evidence that this is also true for
non-local AGN at intermediate redshifts $z \sim 1$. \citet{mendez16} studied the
clustering of AGN in the PRIMUS and DEEP2 surveys (including the COSMOS field)
at $z \sim 0.7$ based on multiple selection criteria. In their X-ray selected
AGN sample, they did not find a significant dependence on clustering in terms of
specific BH accretion rate, in line with our results.

\subsection{Clustering in terms of host galaxy stellar mass}

\begin{figure*}[htbp]
    \resizebox{\hsize}{!}{
        \includegraphics[width=\textwidth]{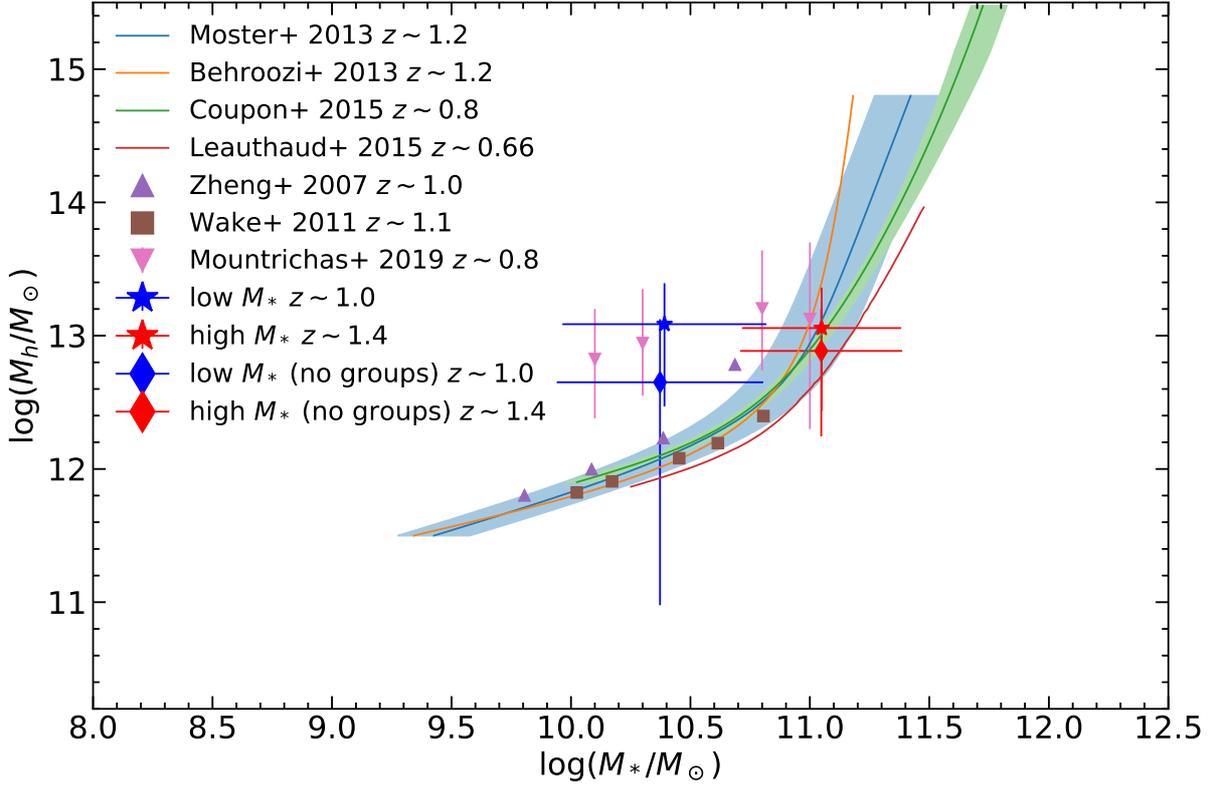}
    }
    \caption{The $M_*-M_\mathrm{halo}$ relationship for our spectroscopic
    redshift AGN sample (stars) compared to previous studies in literature
    according to the legend. For each of the $M_*$ subsamples, the horizontal
    errorbars represent one standard deviation of $\log M_*$ of the sample.
    }
    \label{fig:mstar_mhalo}
\end{figure*}

We also studied the AGN clustering dependence on host galaxy stellar mass,
probing the $M_*-M_\mathrm{halo}$ relation for active galaxies. In Figure
\ref{fig:mstar_mhalo}, we compare our results for \textit{XMM}-COSMOS AGN with
recent studies in literature using normal (non-active) galaxies. For our
comparison, we convert the results to our adopted $h=0.7$ cosmology. DM halo
masses defined with respect to 200 times $\rho_\mathrm{crit}$ have been
re-defined to be with respect to mean density of the background. The blue curve
shows the \citet{moster13} $M_*-M_\mathrm{halo}$ relation for central galaxies
estimated using a multi-epoch abundance matching method which we have calculated
at the mean redshift $z \sim 1.2$ of our AGN sample. The orange curve shows the
galaxy $M_*-M_\mathrm{halo}$ relation of \citet{behroozi13b} at $z \sim 1.2$.
\citet{coupon15} estimated the $M_*-M_\mathrm{halo}$ relation in the
CFHTLenS/VIPERS field at $z \sim 0.8$ using constraints from several different
methods including galaxy clustering. Compared to our AGN sample, their sample
has a similar range in stellar mass and a slightly lower redshift. Results from
HOD modeling of galaxy clustering in DEEP2 \citep{zheng07} and the NMBS
\citep{wake11} at comparable redshifts ($z \sim 1.0-1.1$) are shown as well.
Using weak lensing methods, \citet{leauthaud15} studied a sample of
moderate-luminosity AGN in COSMOS at a lower redshift $z \sim 0.66$ than our
sample. At $M_* > 10^{10.5} \,\mathrm{M}_\sun$, they suggest that AGN populate
similar DM halos as normal galaxies. Similarly, we found that \emph{high} $M_*$
($\gtrsim 10^{10.5} \,\mathrm{M}_\odot$) \emph{XMM}-COSMOS AGN follow the same
$M_*-M_\mathrm{halo}$ relation as normal non-active galaxies. On the contrary,
we estimated that \emph{low} $M_*$ ($\lesssim 10^{10.5} \,\mathrm{M}_\odot$) AGN
are more clustered than normal galaxies. \citet{mountrichas19} measured
clustering of AGN from the \textit{XMM}-XXL survey in terms of host galaxy
properties ($M_*$, SFR, sSFR) at $z \sim 0.8$ and find a positive dependence on
the environment with respect to $M_*$. Within errors, our results at slightly
higher redshift are in agreement with their measurements (see Figure
\ref{fig:mstar_mhalo}).

The $M_*-M_\mathrm{halo}$ relation obtained from our clustering analysis of
\textit{XMM}-COSMOS AGN is not consistent with results inferred for normal
galaxies at similar redshifts, at least for the \textit{low} $M_*$ bin. In fact,
we found that AGN host galaxies with \textit{low} $M_*$ reside in slightly more
massive halos than normal galaxies of similar stellar mass. On the other hand,
at \textit{high} $M_*$, our results are in good agreement with the
$M_*-M_\mathrm{halo}$ relation of normal galaxies. Following Figure
\ref{fig:mstar_mhalo}, we do not expect the observed discrepancy at \textit{low}
$M_*$ to be due to the different mean redshift of the two subsamples ($z \sim 1$
and $z \sim 1.4$). If we exclude AGN that are associated with galaxy groups from
our $M_*$ subsamples, we see that this affects our \textit{low} $M_*$ bin more,
while leaving the \textit{high} $M_*$ bin relatively unchanged. This could
indicate that \textit{XMM}-COSMOS AGN with higher $M_*$ are more preferably
found in central galaxies of their respective halos. For lower $M_*$, the
fraction of AGNs as satellites would be higher. Nevertheless, excluding the
galaxy groups from the analysis brings our result for the \textit{low} $M_*$
closer to the $M_*-M_\mathrm{halo}$ of normal non-active galaxies.

It is important to note that our results for the $M_*$ subsamples include both
type1 and type2 AGNs i.e. AGNs regardless of obscuration are considered in the
same subsample. With the limited sample size of \emph{XMM}-COSMOS, we are not
able to further divide the subsamples and examine the $M_*-M_\mathrm{halo}$
relation for type1 and type2 AGNs separately, to see whether there are any
differences between these to populations. However, this issue can be revisited
with \emph{Chandra} COSMOS Legacy Survey AGNs.

\section{Conclusions}

We have measured the clustering of \textit{XMM}-COSMOS AGN in terms of host
galaxy stellar mass $M_*$ and specific BH accretion rate $L_X/M_*$. Using these
two quantities, we created AGN subsamples by splitting the full sample in terms
of one quantity, while matching the distribution in the other. In addition, we
investigated including AGNs with photometric redshifts as Pdfs in addition to
AGNs with known spectroscopic redshifts. From our analysis, we make the
following conclusions:

\begin{enumerate}

    \item \textit{XMM}-COSMOS AGNs are highly biased
        with a typical DM halo mass of $M_\mathrm{halo} \sim 10^{13}
        \,{h}^{-1}\mathrm{M}_\odot$, characteristic to group-sized environments
        and in broad agreement with previous results for moderate-luminosity
        X-ray selected AGN.

    \item We find no significant clustering dependence in terms of specific BH
        accretion rate, consistent with a picture that higher accretion rates in
        AGNs do not necessarily correspond to more dense environments.

    \item Also we find no significant clustering dependence in terms of host
        galaxy stellar mass. By comparing our results with various
        $M_*-M_\mathrm{halo}$ relations found for normal non-active galaxies,
        we find that our \textit{low} $M_*$ AGN subsample is more clustered than
        what is expected of normal galaxies at similar $M_*$. We investigate
        this further by excluding AGNs that are associated with galaxy groups.
        We find that excluding objects in galaxy groups results in a lower AGN
        bias for the \textit{low} $M_*$ AGN subsamples, but does not affect
        \textit{high} $M_*$. This could be due to a higher fraction of
        satellites for the lower stellar mass systems.

    \item Our selected quality criterion for including additional photometric
        redshifts as Pdfs decreases the errors on the measured best-fit bias and
        does not introduce a bias to the clustering signal. Optimal quality cuts
        for including photometric redshifts will be studied in a future work.

\end{enumerate}

\begin{acknowledgements}

We thank the referee for helpful comments that have improved this paper. AV is
    grateful to the Vilho, Yrjö and Kalle Väisälä Foundation of the Finnish
    Academy of Science and Letters. VA acknowledges funding from the European
    Union's Horizon 2020 research and innovation programme under grant agreement
    No 749348. TM is supported by UNAM-DGAPA PAPIIT IN111379 and CONACyT 252531.
    RG acknowledges support from the agreement ASI-INAF n.2017-14-H.O.
\end{acknowledgements}


\begin{thebibliography}{53}
\expandafter\ifx\csname natexlab\endcsname\relax\def\natexlab#1{#1}\fi

\bibitem[{{Alexander} \& {Hickox}(2012)}]{alexander_hickox12}
{Alexander}, D.~M. \& {Hickox}, R.~C. 2012, \nar, 56, 93

\bibitem[{{Allevato} {et~al.}(2016){Allevato}, {Civano}, {Finoguenov},
  {Marchesi}, {Shankar}, {Zamorani}, {Hasinger}, {Salvato}, {Miyaji}, {Gilli},
  {Cappelluti}, {Brusa}, {Suh}, {Lanzuisi}, {Trakhtenbrot}, {Griffiths},
  {Vignali}, {Schawinski}, \& {Karim}}]{allevato16}
{Allevato}, V., {Civano}, F., {Finoguenov}, A., {et~al.} 2016, \apj, 832, 70

\bibitem[{{Allevato} {et~al.}(2011){Allevato}, {Finoguenov}, {Cappelluti},
  {Miyaji}, {Hasinger}, {Salvato}, {Brusa}, {Gilli}, {Zamorani}, {Shankar},
  {James}, {McCracken}, {Bongiorno}, {Merloni}, {Peacock}, {Silverman}, \&
  {Comastri}}]{allevato11}
{Allevato}, V., {Finoguenov}, A., {Cappelluti}, N., {et~al.} 2011, \apj, 736,
  99

\bibitem[{{Allevato} {et~al.}(2014){Allevato}, {Finoguenov}, {Civano},
  {Cappelluti}, {Shankar}, {Miyaji}, {Hasinger}, {Gilli}, {Zamorani},
  {Lanzuisi}, {Salvato}, {Elvis}, {Comastri}, \& {Silverman}}]{allevato14a}
{Allevato}, V., {Finoguenov}, A., {Civano}, F., {et~al.} 2014, \apj, 796, 4

\bibitem[{{Allevato} {et~al.}(2012){Allevato}, {Finoguenov}, {Hasinger},
  {Miyaji}, {Cappelluti}, {Salvato}, {Zamorani}, {Gilli}, {George}, {Tanaka},
  {Brusa}, {Silverman}, {Civano}, {Elvis}, \& {Shankar}}]{allevato12}
{Allevato}, V., {Finoguenov}, A., {Hasinger}, G., {et~al.} 2012, \apj, 758, 47

\bibitem[{{Behroozi} {et~al.}(2013){Behroozi}, {Wechsler}, \&
  {Conroy}}]{behroozi13b}
{Behroozi}, P.~S., {Wechsler}, R.~H., \& {Conroy}, C. 2013, \apj, 770, 57

\bibitem[{{Bongiorno} {et~al.}(2012){Bongiorno}, {Merloni}, {Brusa},
  {Magnelli}, {Salvato}, {Mignoli}, {Zamorani}, {Fiore}, {Rosario}, {Mainieri},
  {Hao}, {Comastri}, {Vignali}, {Balestra}, {Bardelli}, {Berta}, {Civano},
  {Kampczyk}, {Le Floc'h}, {Lusso}, {Lutz}, {Pozzetti}, {Pozzi}, {Riguccini},
  {Shankar}, \& {Silverman}}]{bongiorno12}
{Bongiorno}, A., {Merloni}, A., {Brusa}, M., {et~al.} 2012, \mnras, 427, 3103

\bibitem[{{Brusa} {et~al.}(2010){Brusa}, {Civano}, {Comastri}, {Miyaji},
  {Salvato}, {Zamorani}, {Cappelluti}, {Fiore}, {Hasinger}, {Mainieri},
  {Merloni}, {Bongiorno}, {Capak}, {Elvis}, {Gilli}, {Hao}, {Jahnke},
  {Koekemoer}, {Ilbert}, {Le Floc'h}, {Lusso}, {Mignoli}, {Schinnerer},
  {Silverman}, {Treister}, {Trump}, {Vignali}, {Zamojski}, {Aldcroft},
  {Aussel}, {Bardelli}, {Bolzonella}, {Cappi}, {Caputi}, {Contini},
  {Finoguenov}, {Fruscione}, {Garilli}, {Impey}, {Iovino}, {Iwasawa},
  {Kampczyk}, {Kartaltepe}, {Kneib}, {Knobel}, {Kovac}, {Lamareille},
  {Leborgne}, {Le Brun}, {Le Fevre}, {Lilly}, {Maier}, {McCracken}, {Pello},
  {Peng}, {Perez-Montero}, {de Ravel}, {Sanders}, {Scodeggio}, {Scoville},
  {Tanaka}, {Taniguchi}, {Tasca}, {de la Torre}, {Tresse}, {Vergani}, \&
  {Zucca}}]{brusa10}
{Brusa}, M., {Civano}, F., {Comastri}, A., {et~al.} 2010, \apj, 716, 348

\bibitem[{{Cappelluti} {et~al.}(2012){Cappelluti}, {Allevato}, \&
  {Finoguenov}}]{cappelluti12}
{Cappelluti}, N., {Allevato}, V., \& {Finoguenov}, A. 2012, Advances in
  Astronomy, 2012, 853701

\bibitem[{{Cappelluti} {et~al.}(2009){Cappelluti}, {Brusa}, {Hasinger},
  {Comastri}, {Zamorani}, {Finoguenov}, {Gilli}, {Puccetti}, {Miyaji},
  {Salvato}, {Vignali}, {Aldcroft}, {B{\"o}hringer}, {Brunner}, {Civano},
  {Elvis}, {Fiore}, {Fruscione}, {Griffiths}, {Guzzo}, {Iovino}, {Koekemoer},
  {Mainieri}, {Scoville}, {Shopbell}, {Silverman}, \& {Urry}}]{cappelluti09}
{Cappelluti}, N., {Brusa}, M., {Hasinger}, G., {et~al.} 2009, \aap, 497, 635

\bibitem[{{Cappelluti} {et~al.}(2007){Cappelluti}, {Hasinger}, {Brusa},
  {Comastri}, {Zamorani}, {B{\"o}hringer}, {Brunner}, {Civano}, {Finoguenov},
  {Fiore}, {Gilli}, {Griffiths}, {Mainieri}, {Matute}, {Miyaji}, \&
  {Silverman}}]{cappelluti07}
{Cappelluti}, N., {Hasinger}, G., {Brusa}, M., {et~al.} 2007, \apjs, 172, 341

\bibitem[{{Civano} {et~al.}(2016){Civano}, {Marchesi}, {Comastri}, {Urry},
  {Elvis}, {Cappelluti}, {Puccetti}, {Brusa}, {Zamorani}, {Hasinger},
  {Aldcroft}, {Alexander}, {Allevato}, {Brunner}, {Capak}, {Finoguenov},
  {Fiore}, {Fruscione}, {Gilli}, {Glotfelty}, {Griffiths}, {Hao}, {Harrison},
  {Jahnke}, {Kartaltepe}, {Karim}, {LaMassa}, {Lanzuisi}, {Miyaji}, {Ranalli},
  {Salvato}, {Sargent}, {Scoville}, {Schawinski}, {Schinnerer}, {Silverman},
  {Smolcic}, {Stern}, {Toft}, {Trakhtenbrot}, {Treister}, \&
  {Vignali}}]{civano16}
{Civano}, F., {Marchesi}, S., {Comastri}, A., {et~al.} 2016, \apj, 819, 62

\bibitem[{{Coil} {et~al.}(2009){Coil}, {Georgakakis}, {Newman}, {Cooper},
  {Croton}, {Davis}, {Koo}, {Laird}, {Nandra}, {Weiner}, {Willmer}, \&
  {Yan}}]{coil09}
{Coil}, A.~L., {Georgakakis}, A., {Newman}, J.~A., {et~al.} 2009, \apj, 701,
  1484

\bibitem[{{Cooray} \& {Sheth}(2002)}]{cooray_sheth02}
{Cooray}, A. \& {Sheth}, R. 2002, \physrep, 372, 1

\bibitem[{{Coupon} {et~al.}(2015){Coupon}, {Arnouts}, {van Waerbeke},
  {Moutard}, {Ilbert}, {van Uitert}, {Erben}, {Garilli}, {Guzzo}, {Heymans},
  {Hildebrandt}, {Hoekstra}, {Kilbinger}, {Kitching}, {Mellier}, {Miller},
  {Scodeggio}, {Bonnett}, {Branchini}, {Davidzon}, {De Lucia}, {Fritz}, {Fu},
  {Hudelot}, {Hudson}, {Kuijken}, {Leauthaud}, {Le F{\`e}vre}, {McCracken},
  {Moscardini}, {Rowe}, {Schrabback}, {Semboloni}, \& {Velander}}]{coupon15}
{Coupon}, J., {Arnouts}, S., {van Waerbeke}, L., {et~al.} 2015, \mnras, 449,
  1352

\bibitem[{{Croom} {et~al.}(2005){Croom}, {Boyle}, {Shanks}, {Smith}, {Miller},
  {Outram}, {Loaring}, {Hoyle}, \& {da {\^A}ngela}}]{croom05}
{Croom}, S.~M., {Boyle}, B.~J., {Shanks}, T., {et~al.} 2005, \mnras, 356, 415

\bibitem[{{da {\^A}ngela} {et~al.}(2008){da {\^A}ngela}, {Shanks}, {Croom},
  {Weilbacher}, {Brunner}, {Couch}, {Miller}, {Myers}, {Nichol}, {Pimbblet},
  {de Propris}, {Richards}, {Ross}, {Schneider}, \& {Wake}}]{daangela08}
{da {\^A}ngela}, J., {Shanks}, T., {Croom}, S.~M., {et~al.} 2008, \mnras, 383,
  565

\bibitem[{{Davis} \& {Peebles}(1983)}]{davis_peebles83}
{Davis}, M. \& {Peebles}, P.~J.~E. 1983, \apj, 267, 465

\bibitem[{{Eisenstein} \& {Hu}(1999)}]{eisenstein_hu99}
{Eisenstein}, D.~J. \& {Hu}, W. 1999, \apj, 511, 5

\bibitem[{{Fanidakis} {et~al.}(2013){Fanidakis}, {Georgakakis}, {Mountrichas},
  {Krumpe}, {Baugh}, {Lacey}, {Frenk}, {Miyaji}, \& {Benson}}]{fanidakis13}
{Fanidakis}, N., {Georgakakis}, A., {Mountrichas}, G., {et~al.} 2013, \mnras,
  435, 679

\bibitem[{{Finoguenov} {et~al.}(2007){Finoguenov}, {Guzzo}, {Hasinger},
  {Scoville}, {Aussel}, {B{\"o}hringer}, {Brusa}, {Capak}, {Cappelluti},
  {Comastri}, {Giodini}, {Griffiths}, {Impey}, {Koekemoer}, {Kneib},
  {Leauthaud}, {Le F{\`e}vre}, {Lilly}, {Mainieri}, {Massey}, {McCracken},
  {Mobasher}, {Murayama}, {Peacock}, {Sakelliou}, {Schinnerer}, {Silverman},
  {Smol{\v c}i{\'c}}, {Taniguchi}, {Tasca}, {Taylor}, {Trump}, \&
  {Zamorani}}]{finoguenov07}
{Finoguenov}, A., {Guzzo}, L., {Hasinger}, G., {et~al.} 2007, \apjs, 172, 182

\bibitem[{{Georgakakis} {et~al.}(2014){Georgakakis}, {Mountrichas}, {Salvato},
  {Rosario}, {P{\'e}rez-Gonz{\'a}lez}, {Lutz}, {Nandra}, {Coil}, {Cooper},
  {Newman}, {Berta}, {Magnelli}, {Popesso}, \& {Pozzi}}]{georgakakis14}
{Georgakakis}, A., {Mountrichas}, G., {Salvato}, M., {et~al.} 2014, \mnras,
  443, 3327

\bibitem[{{George} {et~al.}(2011){George}, {Leauthaud}, {Bundy}, {Finoguenov},
  {Tinker}, {Lin}, {Mei}, {Kneib}, {Aussel}, {Behroozi}, {Busha}, {Capak},
  {Coccato}, {Covone}, {Faure}, {Fiorenza}, {Ilbert}, {Le Floc'h}, {Koekemoer},
  {Tanaka}, {Wechsler}, \& {Wolk}}]{george11}
{George}, M.~R., {Leauthaud}, A., {Bundy}, K., {et~al.} 2011, \apj, 742, 125

\bibitem[{{Gilli} {et~al.}(2009){Gilli}, {Zamorani}, {Miyaji}, {Silverman},
  {Brusa}, {Mainieri}, {Cappelluti}, {Daddi}, {Porciani}, {Pozzetti}, {Civano},
  {Comastri}, {Finoguenov}, {Fiore}, {Salvato}, {Vignali}, {Hasinger}, {Lilly},
  {Impey}, {Trump}, {Capak}, {McCracken}, {Scoville}, {Taniguchi}, {Carollo},
  {Contini}, {Kneib}, {Le Fevre}, {Renzini}, {Scodeggio}, {Bardelli},
  {Bolzonella}, {Bongiorno}, {Caputi}, {Cimatti}, {Coppa}, {Cucciati}, {de La
  Torre}, {de Ravel}, {Franzetti}, {Garilli}, {Iovino}, {Kampczyk}, {Knobel},
  {Kova{\v c}}, {Lamareille}, {Le Borgne}, {Le Brun}, {Maier}, {Mignoli},
  {Pell{\`o}}, {Peng}, {Perez Montero}, {Ricciardelli}, {Tanaka}, {Tasca},
  {Tresse}, {Vergani}, {Zucca}, {Abbas}, {Bottini}, {Cappi}, {Cassata},
  {Fumana}, {Guzzo}, {Leauthaud}, {Maccagni}, {Marinoni}, {Memeo}, {Meneux},
  {Oesch}, {Scaramella}, \& {Walcher}}]{gilli09}
{Gilli}, R., {Zamorani}, G., {Miyaji}, T., {et~al.} 2009, \aap, 494, 33

\bibitem[{{Hasinger} {et~al.}(2018){Hasinger}, {Capak}, {Salvato}, {Barger},
  {Cowie}, {Faisst}, {Hemmati}, {Kakazu}, {Kartaltepe}, {Masters}, {Mobasher},
  {Nayyeri}, {Sanders}, {Scoville}, {Suh}, {Steinhardt}, \&
  {Yang}}]{hasinger18}
{Hasinger}, G., {Capak}, P., {Salvato}, M., {et~al.} 2018, \apj, 858, 77

\bibitem[{{Hasinger} {et~al.}(2007){Hasinger}, {Cappelluti}, {Brunner},
  {Brusa}, {Comastri}, {Elvis}, {Finoguenov}, {Fiore}, {Franceschini}, {Gilli},
  {Griffiths}, {Lehmann}, {Mainieri}, {Matt}, {Matute}, {Miyaji}, {Molendi},
  {Paltani}, {Sanders}, {Scoville}, {Tresse}, {Urry}, {Vettolani}, \&
  {Zamorani}}]{hasinger07}
{Hasinger}, G., {Cappelluti}, N., {Brunner}, H., {et~al.} 2007, \apjs, 172, 29

\bibitem[{{Hickox} {et~al.}(2009){Hickox}, {Jones}, {Forman}, {Murray},
  {Kochanek}, {Eisenstein}, {Jannuzi}, {Dey}, {Brown}, {Stern}, {Eisenhardt},
  {Gorjian}, {Brodwin}, {Narayan}, {Cool}, {Kenter}, {Caldwell}, \&
  {Anderson}}]{hickox09a}
{Hickox}, R.~C., {Jones}, C., {Forman}, W.~R., {et~al.} 2009, \apj, 696, 891

\bibitem[{{Koutoulidis} {et~al.}(2018){Koutoulidis}, {Georgantopoulos},
  {Mountrichas}, {Plionis}, {Georgakakis}, {Akylas}, \&
  {Rovilos}}]{koutoulidis18}
{Koutoulidis}, L., {Georgantopoulos}, I., {Mountrichas}, G., {et~al.} 2018,
  \mnras, 481, 3063

\bibitem[{{Koutoulidis} {et~al.}(2013){Koutoulidis}, {Plionis},
  {Georgantopoulos}, \& {Fanidakis}}]{koutoulidis13}
{Koutoulidis}, L., {Plionis}, M., {Georgantopoulos}, I., \& {Fanidakis}, N.
  2013, \mnras, 428, 1382

\bibitem[{{Krumpe} {et~al.}(2014){Krumpe}, {Miyaji}, \& {Coil}}]{krumpe14}
{Krumpe}, M., {Miyaji}, T., \& {Coil}, A.~L. 2014, in Multifrequency Behaviour
  of High Energy Cosmic Sources, 71--78

\bibitem[{{Krumpe} {et~al.}(2018){Krumpe}, {Miyaji}, {Coil}, \&
  {Aceves}}]{krumpe18}
{Krumpe}, M., {Miyaji}, T., {Coil}, A.~L., \& {Aceves}, H. 2018, \mnras, 474,
  1773

\bibitem[{{Krumpe} {et~al.}(2015){Krumpe}, {Miyaji}, {Husemann}, {Fanidakis},
  {Coil}, \& {Aceves}}]{krumpe15}
{Krumpe}, M., {Miyaji}, T., {Husemann}, B., {et~al.} 2015, \apj, 815, 21

\bibitem[{{Landy} \& {Szalay}(1993)}]{landy_szalay93}
{Landy}, S.~D. \& {Szalay}, A.~S. 1993, \apj, 412, 64

\bibitem[{{Leauthaud} {et~al.}(2010){Leauthaud}, {Finoguenov}, {Kneib},
  {Taylor}, {Massey}, {Rhodes}, {Ilbert}, {Bundy}, {Tinker}, {George}, {Capak},
  {Koekemoer}, {Johnston}, {Zhang}, {Cappelluti}, {Ellis}, {Elvis}, {Giodini},
  {Heymans}, {Le F{\`e}vre}, {Lilly}, {McCracken}, {Mellier},
  {R{\'e}fr{\'e}gier}, {Salvato}, {Scoville}, {Smoot}, {Tanaka}, {Van
  Waerbeke}, \& {Wolk}}]{leauthaud10}
{Leauthaud}, A., {Finoguenov}, A., {Kneib}, J.-P., {et~al.} 2010, \apj, 709, 97

\bibitem[{{Leauthaud} {et~al.}(2015){Leauthaud}, {J.~Benson}, {Civano},
  {L.~Coil}, {Bundy}, {Massey}, {Schramm}, {Schulze}, {Capak}, {Elvis},
  {Kulier}, \& {Rhodes}}]{leauthaud15}
{Leauthaud}, A., {J.~Benson}, A., {Civano}, F., {et~al.} 2015, \mnras, 446,
  1874

\bibitem[{{Marchesi} {et~al.}(2016){Marchesi}, {Civano}, {Elvis}, {Salvato},
  {Brusa}, {Comastri}, {Gilli}, {Hasinger}, {Lanzuisi}, {Miyaji}, {Treister},
  {Urry}, {Vignali}, {Zamorani}, {Allevato}, {Cappelluti}, {Cardamone},
  {Finoguenov}, {Griffiths}, {Karim}, {Laigle}, {LaMassa}, {Jahnke}, {Ranalli},
  {Schawinski}, {Schinnerer}, {Silverman}, {Smolcic}, {Suh}, \&
  {Trakhtenbrot}}]{marchesi16}
{Marchesi}, S., {Civano}, F., {Elvis}, M., {et~al.} 2016, \apj, 817, 34

\bibitem[{{Marulli} {et~al.}(2016){Marulli}, {Veropalumbo}, \&
  {Moresco}}]{marulli16}
{Marulli}, F., {Veropalumbo}, A., \& {Moresco}, M. 2016, Astronomy and
  Computing, 14, 35

\bibitem[{{Mendez} {et~al.}(2016){Mendez}, {Coil}, {Aird}, {Skibba},
  {Diamond-Stanic}, {Moustakas}, {Blanton}, {Cool}, {Eisenstein}, {Wong}, \&
  {Zhu}}]{mendez16}
{Mendez}, A.~J., {Coil}, A.~L., {Aird}, J., {et~al.} 2016, \apj, 821, 55

\bibitem[{{Merloni} {et~al.}(2019){Merloni}, {Alexander}, {Banerji}, {Boller},
  {Comparat}, {Dwelly}, {Fotopoulou}, {McMahon}, {Nandra}, {Salvato}, {Croom},
  {Finoguenov}, {Krumpe}, {Lamer}, {Rosario}, {Schwope}, {Shanks}, {Steinmetz},
  {Wisotzki}, \& {Worseck}}]{merloni19}
{Merloni}, A., {Alexander}, D.~A., {Banerji}, M., {et~al.} 2019, The Messenger,
  175, 42

\bibitem[{{Miyaji} {et~al.}(2007){Miyaji}, {Zamorani}, {Cappelluti}, {Gilli},
  {Griffiths}, {Comastri}, {Hasinger}, {Brusa}, {Fiore}, {Puccetti}, {Guzzo},
  \& {Finoguenov}}]{miyaji07}
{Miyaji}, T., {Zamorani}, G., {Cappelluti}, N., {et~al.} 2007, \apjs, 172, 396

\bibitem[{{Moster} {et~al.}(2013){Moster}, {Naab}, \& {White}}]{moster13}
{Moster}, B.~P., {Naab}, T., \& {White}, S.~D.~M. 2013, \mnras, 428, 3121

\bibitem[{{Mountrichas} {et~al.}(2019){Mountrichas}, {Georgakakis}, \&
  {Georgantopoulos}}]{mountrichas19}
{Mountrichas}, G., {Georgakakis}, A., \& {Georgantopoulos}, I. 2019, \mnras,
  483, 1374

\bibitem[{{Mountrichas} {et~al.}(2016){Mountrichas}, {Georgakakis}, {Menzel},
  {Fanidakis}, {Merloni}, {Liu}, {Salvato}, \& {Nandra}}]{mountrichas16}
{Mountrichas}, G., {Georgakakis}, A., {Menzel}, M.-L., {et~al.} 2016, \mnras,
  457, 4195

\bibitem[{{Norberg} {et~al.}(2009){Norberg}, {Baugh}, {Gazta{\~n}aga}, \&
  {Croton}}]{norberg09}
{Norberg}, P., {Baugh}, C.~M., {Gazta{\~n}aga}, E., \& {Croton}, D.~J. 2009,
  \mnras, 396, 19

\bibitem[{{Powell} {et~al.}(2018){Powell}, {Cappelluti}, {Urry}, {Koss},
  {Finoguenov}, {Ricci}, {Trakhtenbrot}, {Allevato}, {Ajello}, {Oh},
  {Schawinski}, \& {Secrest}}]{powell18}
{Powell}, M.~C., {Cappelluti}, N., {Urry}, C.~M., {et~al.} 2018, \apj, 858, 110

\bibitem[{{Ross} {et~al.}(2009){Ross}, {Shen}, {Strauss}, {Vanden Berk},
  {Connolly}, {Richards}, {Schneider}, {Weinberg}, {Hall}, {Bahcall}, \&
  {Brunner}}]{ross09}
{Ross}, N.~P., {Shen}, Y., {Strauss}, M.~A., {et~al.} 2009, \apj, 697, 1634

\bibitem[{{Salvato} {et~al.}(2009){Salvato}, {Hasinger}, {Ilbert}, {Zamorani},
  {Brusa}, {Scoville}, {Rau}, {Capak}, {Arnouts}, {Aussel}, {Bolzonella},
  {Buongiorno}, {Cappelluti}, {Caputi}, {Civano}, {Cook}, {Elvis}, {Gilli},
  {Jahnke}, {Kartaltepe}, {Impey}, {Lamareille}, {Le Floc'h}, {Lilly},
  {Mainieri}, {McCarthy}, {McCracken}, {Mignoli}, {Mobasher}, {Murayama},
  {Sasaki}, {Sanders}, {Schiminovich}, {Shioya}, {Shopbell}, {Silverman},
  {Smol{\v c}i{\'c}}, {Surace}, {Taniguchi}, {Thompson}, {Trump}, {Urry}, \&
  {Zamojski}}]{salvato09}
{Salvato}, M., {Hasinger}, G., {Ilbert}, O., {et~al.} 2009, \apj, 690, 1250

\bibitem[{{Salvato} {et~al.}(2011){Salvato}, {Ilbert}, {Hasinger}, {Rau},
  {Civano}, {Zamorani}, {Brusa}, {Elvis}, {Vignali}, {Aussel}, {Comastri},
  {Fiore}, {Le Floc'h}, {Mainieri}, {Bardelli}, {Bolzonella}, {Bongiorno},
  {Capak}, {Caputi}, {Cappelluti}, {Carollo}, {Contini}, {Garilli}, {Iovino},
  {Fotopoulou}, {Fruscione}, {Gilli}, {Halliday}, {Kneib}, {Kakazu},
  {Kartaltepe}, {Koekemoer}, {Kovac}, {Ideue}, {Ikeda}, {Impey}, {Le Fevre},
  {Lamareille}, {Lanzuisi}, {Le Borgne}, {Le Brun}, {Lilly}, {Maier},
  {Manohar}, {Masters}, {McCracken}, {Messias}, {Mignoli}, {Mobasher}, {Nagao},
  {Pello}, {Puccetti}, {Perez-Montero}, {Renzini}, {Sargent}, {Sanders},
  {Scodeggio}, {Scoville}, {Shopbell}, {Silvermann}, {Taniguchi}, {Tasca},
  {Tresse}, {Trump}, \& {Zucca}}]{salvato11}
{Salvato}, M., {Ilbert}, O., {Hasinger}, G., {et~al.} 2011, \apj, 742, 61

\bibitem[{{Scoville} {et~al.}(2007){Scoville}, {Aussel}, {Brusa}, {Capak},
  {Carollo}, {Elvis}, {Giavalisco}, {Guzzo}, {Hasinger}, {Impey}, {Kneib},
  {LeFevre}, {Lilly}, {Mobasher}, {Renzini}, {Rich}, {Sanders}, {Schinnerer},
  {Schminovich}, {Shopbell}, {Taniguchi}, \& {Tyson}}]{scoville07}
{Scoville}, N., {Aussel}, H., {Brusa}, M., {et~al.} 2007, \apjs, 172, 1

\bibitem[{{Sheth} {et~al.}(2001){Sheth}, {Mo}, \& {Tormen}}]{sheth01}
{Sheth}, R.~K., {Mo}, H.~J., \& {Tormen}, G. 2001, \mnras, 323, 1

\bibitem[{{van den Bosch}(2002)}]{vandenbosch02}
{van den Bosch}, F.~C. 2002, \mnras, 331, 98

\bibitem[{{Wake} {et~al.}(2011){Wake}, {Whitaker}, {Labb{\'e}}, {van Dokkum},
  {Franx}, {Quadri}, {Brammer}, {Kriek}, {Lundgren}, {Marchesini}, \&
  {Muzzin}}]{wake11}
{Wake}, D.~A., {Whitaker}, K.~E., {Labb{\'e}}, I., {et~al.} 2011, \apj, 728, 46

\bibitem[{{Zheng} {et~al.}(2007){Zheng}, {Coil}, \& {Zehavi}}]{zheng07}
{Zheng}, Z., {Coil}, A.~L., \& {Zehavi}, I. 2007, \apj, 667, 760

\end{thebibliography}
\end{document}